\documentclass[12pt,letterpaper]{article}

\usepackage{amsmath}
\usepackage{amsfonts}
\usepackage{amssymb}
\usepackage{graphicx}

\def\ee{\end{eqnarray}}

\def\d{\partial}
\def\nn{\nonumber}
\newcommand{\dd}[2]{\frac{\partial #1}{ \partial #2}}

\newcommand{\half}{{\textstyle{\frac12}}}
\newcommand{\di}{\partial}

\newcommand{\be}{\begin{eqnarray}}
\newcommand{\en}{\end{eqnarray}}
\newcommand{\bea}[1]{\left(\begin{array}{#1}}
\newcommand{\ena}{\end{array}\right)}

\begin{document}

\vspace{5mm}
\vspace{0.5cm}
\begin{center}

\def\thefootnote{\fnsymbol{footnote}}

{\Large \bf On the consistency relation of the 3-point function
\\in single field inflation}
\\[0.5cm]
{\large Clifford Cheung, A.~Liam Fitzpatrick,\\[.1cm]
Jared Kaplan
and Leonardo Senatore}
\\[0.5cm]

{\small \textit{
Jefferson Physical Laboratory \\
Harvard University, Cambridge, MA 02138, USA}}



\end{center}

\vspace{.8cm}

\hrule \vspace{0.3cm}
{\small  \noindent \textbf{Abstract} \\[0.3cm]
\noindent
The consistency relation for the 3-point function of the CMB is a very
powerful observational signature which is believed to be true for every inflationary model in which
there is
only one dynamical degree of freedom. Its importance relies on
the fact that deviations from it might be detected in next generation experiments,
allowing us to rule out all single field inflationary models.
After making more precise the already existing proof of the consistency relation, we use a
recently developed effective field theory for inflationary perturbations to provide an
alternative and very explicit proof valid at leading non trivial order
in slow roll parameters.
\vspace{0.5cm}  \hrule
\def\thefootnote{\arabic{footnote}}
\setcounter{footnote}{0}

\section{Introduction}

In the last few years there has been great progress in
understanding the non-Gaussianity of the primordial spectrum
of density fluctuations. Starting from Maldacena's first full
computation of the non-Gaussian features in single field slow roll
inflation \cite{Maldacena:2002vr}, several alternative models
have been proposed that produce a large and in principle detectable
level of non-Gaussianities
\cite{Arkani-Hamed:2003uz,Alishahiha:2004eh,Chen:2006nt,Zaldarriaga:2003my}
through different mechanisms for generating density
fluctuations in the quasi de Sitter inflationary phase. At the same
time, from the experimental side, the WMAP satellite has allowed
for a huge improvement in our measurement of the properties
of the CMB. Observations seem to confirm the generic predictions
of standard slow roll inflation \cite{Spergel:2006hy}. Limits on the
primordial non-Gaussianity of the CMB have been significantly improved
\cite{Creminelli:2006rz},
but for the moment the data are consistent with a non-Gaussian signal.

The fact that the CMB seems to be rather Gaussian means that the non-Gaussian
component must be rather small. This makes it clear that the
most important observable for non-Gaussianities will
be the 3-point function of density perturbations \cite{Creminelli:2006gc}
\begin{equation}
\langle\zeta_{\vec{k}_1}\zeta_{\vec{k}_2}\zeta_{\vec{k}_3}\rangle
\end{equation}
where $\zeta_{\vec{k}_1}$ is the density fluctuation of comoving slices in Fourier space.

As pointed out in \cite{Babich:2004gb}, due to symmetry reasons, the
3-point function is a real function of {\it two} variables. While on one
hand this means that it contains a lot of information about the inflationary
model, on the other hand this also means that there really could be a large
number of different shapes of the 3-point function. For this reason,
model independent characterization of the 3-point function are very useful.

To this end, it was pointed out in
\cite{Maldacena:2002vr,Creminelli:2004yq} that in all cases in which there is
only one dynamical field that is important during inflation,
the three point function is connected
to the two point function and its deviation from scale invariance in a particular geometrical limit . In
other words, we have:
\begin{equation}\label{consistency}
\lim_{k_1 \to 0} \langle\zeta_{\vec k_1} \zeta_{\vec k_2} \zeta_{\vec
  k_3}\rangle = - (2 \pi)^3 \delta^3(\sum_i \vec k_i) P_{k_1} P_{k_3}
\frac{d \log k_3^3 P_{k_3}}{d \log k_3} \;,
\end{equation}
where
\begin{equation}\label{two-point}
\langle\zeta_{\vec k_i}\zeta_{\vec k_j}\rangle = (2 \pi)^3 \delta^3(\vec k_i + \vec k_j) P_{k_i} \;.
\end{equation}
Let us comment briefly on what this relationship means. Because of
translation invariance, the spatial momenta $\vec{k}_1,\vec{k}_2,\vec{k}_3$
must form a closed triangle. The consistency relation says that in the limit
in which one of the sides of this triangle (say $\vec{k}_1$) goes to zero --
and therefore the other two sides become
equal and opposite ($\vec{k}_1\simeq- \vec{k}_2$), and the triangle becomes squeezed --
the three point function becomes proportional to the two point function
of the long wavelength modes times the two point function of the short wavelength
mode times its deviation from scale invariance. There are no free parameters: in this
limit the three point function can be fully specified in terms of the two point function.
Notice also that, as we will explicitly show later,
the consistency relation holds without any slow roll approximation. We emphasize 
that our only assumption
 is that there is only one relevant single clock field during the cosmological history.

Experimental limits on non-Guassianities are generically given in terms
of a scalar variable $f_{\rm NL}$ \cite{Komatsu:2003fd, Babich:2004gb} which gives
an amplitude of the 3-point function in the squeezed limit of the form:
\begin{equation}
\lim_{k_1 \to 0} \langle\zeta_{\vec k_1} \zeta_{\vec k_2} \zeta_{\vec
  k_3}\rangle =(2\pi)^3 \delta^3(\sum_i \vec{k}_i)\; 4 f_{\rm NL} P_{k_1} P_{k_3}
\end{equation}
In the limit of small deviation from de Sitter, the consistency relation
predicts that the level of $f_{\rm NL}$ in the squeezed configuration should be
of order of the slow roll parameters ${{\cal O}}(1/100)$. Current limits
from WMAP 3-year data give an $f_{NL}\lesssim 100$ \cite{Creminelli:2006rz},
and the Planck satellite is expected to constrain $f_{\rm NL}$ at the level of a few. We
therefore realize that it is probably impossible, at least in the foreseeable
future, to experimentally verify the consistency relation.

This does not mean that the consistency relation is useless. It is instead
a very powerful instrument in the opposite
regime in which it is proven to be experimentally not satisfied. 
{\it No} model of inflation with only
one dynamical degree of freedom (which therefore acts as a physical 'clock' for
exiting the quasi de Sitter phase) can predict a non-Gaussian signal in the 
squeezed limit of detectable level, meaning that a detection of a signal in 
this limit would allow us to {\it rule out} all single field 
inflationary models.

The assumption of a single 'clock' is really essential, as we will explain, and
in fact there are inflationary models with more than one field that violate 
the consistency relation,
and predict a detectable level on non-Gaussianity in the squeezed limit
(see for example \cite{Lyth:2002my,Zaldarriaga:2003my}). Outside of the inflationay paradigm, the recently proposed new bouncing cosmology
\cite{Lehners:2007ac,Buchbinder:2007ad,Creminelli:2007aq}, though less compelling than inflation, predicts a potentially detectable non-Gaussian signal that violates the consistency relation \cite{Creminelli:2007aq}.
All of this is somewhat reassuring
from the theoretical point of view, as it tells us that we do have some hope
of detecting a deviation from the consistency relation in the near future.

The available arguments that prove the consistency relation
\cite{Maldacena:2002vr,Creminelli:2004yq} are rigorous, but suffer from
 not being very explicit, casting some doubt
on their validity whenever some explicit calculation seems to show that it is
violated.
The purpose of the present paper is to explicitly verify the consistency
relation for all single clock models at leading non trivial order in
slow roll parameters. We can do this because we can exploit
a recently developed effective theory \cite{Creminelli:2006xe, us}
which describes fluctuations around a FRW cosmology for every
single field model. The assumptions
behind the
validity of this effective theory are exactly the same for which the
consistency relation applies, and this allows us to verify the consistency
relation in full generality. Even if general proofs do exist \cite{Maldacena:2002vr,Creminelli:2004yq},
we consider
such an explicit verification worthwhile for the outlined importance of
the consistency relation as being able to rule out all single field models of
inflation.

The paper is structured as follows. In sec. \ref{sec: formal proof} we review and further formalize
 the generic proof of the consistency relation. In particular we highlight
the fact that it is valid beyond any slow roll approximation. In sec.
\ref{sec: full verification} we begin
the verification of the consistency relation at leading order in slow
roll parameters after having briefly
introduced the effective theory we will use. We first study a very large
class of
models which includes for example all the models with a Lagrangian for a
single scalar field of the form $P(-(\partial \phi)^2,\phi)$, explicitly
verifying the consistency relation at first order in slow roll parameters.
Then we go on to
study some more 'exotic' inflationary models that appear in our effective theory
within some simplified assumption. In sec. \ref{sec:conclusions} we summarize and conclude.

\section[]{\label{sec: formal proof} Formal proof of the consistency relation \footnote{The results of this section are obtained in collaboration with Paolo Creminelli.}  }

For simplicity
we concentrate on only the scalar fluctuations and neglect tensor modes \footnote{We make this assumption for simplicity's sake and also because the 3-point function of scalar fluctuations is at least for the moment the most important from the observational point of view. This assumption is not necessary  
for the proof of the consistency relation, as it was previously noted in  
\cite{Maldacena:2002vr,Lyth:2004gb}, and more general fluctuations can be easily included.}.
Using the $\zeta$ variable to describe scalar fluctuations, the
metric for fluctuations around a FRW universe takes the form:
\begin{equation}\label{zetagaugetemp}
ds^2= -N^2 dt^2 + \hat g_{ij} (dx^i + N^i dt) (dx^j + N^j dt) \; ,
\end{equation}
with
\begin{equation}
\hat g_{ij}=a(t)^2 e^{2\zeta}\delta_{ij} \ , \label{zetagauge}
\end{equation}
where we have used the ADM parameterization. In this gauge the
matter is taken to be unperturbed,  fixing the time
diffeomorphisms in this way.

Let us introduce two facts that will be proven to be true later. First, when a mode goes outside the horizon ($\omega\ll H$),
if there is only one degree of freedom, the ADM variable $N$ and
$N^i$ defined in the gauge of (\ref{zetagaugetemp}), go to their unperturbed 
values (respectively 1 and 0)
. In  this limit
the metric becomes
\begin{equation}\label{zetalarge}
ds^2=-dt^2+a(t)^2 e^{2\zeta(x)} dx_i dx_i
\end{equation}
with $\zeta$ constant in time.
Since in this limit we can neglect the gradient terms,
we can re-absorb the $\zeta$ fluctuation in a 'local' rescaling of the coordinates
$x'=e^{\zeta(x)} x$. In regions of space separated by large distances,
the metric becomes the one of unperturbed FRW universes, each one characterized by a
different rescaling of the coordinate (or equivalently of the scale factor). This is sometimes
called the parallel universe description of the inflationary perturbation outside of the
horizon. 
There is a quite intuitive reason why on large scales the metric takes the form 
(\ref{zetalarge}). Once we can neglect gradients, each region of the universe evolves exactly
in the same way, as the inflationary solution is an attractor. As a consequence, the metric
on large scales has to approach the one of the unperturbed FRW universe. The only
difference among the well separated 
observers is how much each region has expanded with respect to the other, and it
is this difference that remains constant. 

The second fact that is important for us is that once a mode goes
outside the horizon, it becomes classical, in the sense that
$ [\zeta_{\vec{k}} , \dot{\zeta}_{\vec{k}'}]\rightarrow 0$
exponentially fast. So for measurements which only involve $\zeta$ or $\dot\zeta$ we can
treat the mode as a classical variable. Proof of this with different approaches can be found for example in \cite{Starobinsky:1982ee,Lyth:2006qz}. 

Let us now concentrate on the three point function of eq. (\ref{consistency}).
We can imagine going to the limit where all the three modes are well outside the horizon,
 so that
we can treat them as classical. We are interested in the regime where $k_1\ll k_2,k_3$.
$\zeta_{\vec{k}_1}$ is therefore a background mode for $\zeta_{\vec{k}_2}$ and
$\zeta_{\vec{k}_3}$. One can therefore compute the three point function in a two
step process: first compute the two point function of $\zeta_{\vec{k_2}}$ and $\zeta_{\vec{k_3}}$
in a background $\zeta^B$:
\begin{equation}
\langle\zeta_{\vec{k}_2}\zeta_{\vec{k}_3}\rangle_{\zeta^B} \ ,
\end{equation}
and then correlate this result with the value of the background field
$\zeta_{\vec{k}_1}$. It is useful to compute this two point function
in real space and so compute
$\langle\zeta_{\vec{x}_2}\zeta_{\vec{x}_3}\rangle$ on the background $\zeta^B(\vec{x})$.
The scale of variation of the background is much larger than $|\vec{x}_2-\vec{x}_3|$.
From this point of view what we are computing is the short scale average on a given
realization of the background. Then, we will correlate with the background and average over
it. Expanding the short scale two point function in powers of the background, we obtain:
\begin{equation}
\label{eq:onaback}
\langle\zeta \zeta\rangle_B (\vec x_2, \vec x_3) \simeq
\langle\zeta\zeta\rangle_0(|\vec x_2- \vec x_3|)+\zeta^B(\frac{\vec x_2+ \vec x_3}{2})
\left.\left(\frac{d}{d\zeta^B}\langle\zeta\zeta\rangle_B(|\vec x_2- \vec x_3|)\right)\right|_0\ ,
\end{equation}
where the subscript $0$ means that the quantity is evaluated on the
vacuum, i.e. without the background wave. 
On the background the two point function does not depend only
on the distance between the two points, but also on their position on
the background. Since the points are very close with respect to the
the typical variation length of the background we can evaluate the
background at the middle point $(\vec x_2 + \vec x_3)/2$. Corrections
to this are sub-leading in the squeezed limit expansion. Notice that no
slow roll approximation has been done: we have just expanded in powers
of the small background field. The background modulates the amplitude
of the two point function; as $\zeta$ is equivalent to a rescaling of
the spatial coordinates, we can trade the derivative with respect to
$\zeta^B$ for a derivative with respect to the log-distance between
the points:
\begin{equation}
\label{eq:onaback2}
\langle\zeta \zeta\rangle_B (\vec x_2, \vec x_3) \simeq
\langle\zeta\zeta\rangle_0(|\vec x_2- \vec x_3|)+\int \frac{d^3k}{(2\pi)^3}\; \zeta^B(\vec{k})
e^{i \vec{k}\cdot (\vec{x}_2+\vec x_3)/2}
\frac{d}{d\log(|\vec x_2-\vec x_3|)}\langle\zeta\zeta\rangle_0(|\vec x_2- \vec x_3|)\ ,
\end{equation}
where we have written $\zeta^B$ in Fourier space:
\begin{equation}
\zeta^B(\vec{x})=\int \frac{d^3k}{(2\pi)^3}\zeta^B(\vec{k}) e^{i \vec{k}\cdot \vec{x}} \ ,
\end{equation}
because this will soon be useful \footnote{Notice that this is our convention
for the Fourier transform.}. Now, we can do the Fourier transform with respect
to $\vec x_2$ and $\vec x_3$. The result can be expressed in terms of $\vec k_L=\vec k_2
+\vec k_3$ and $\vec{k}_S=(\vec k_2-\vec k_3)/2$. Here $_L$ and $_S$ stand for long and short
wavelength. The derivative with respect the log-distance can be integrated by parts to
obtain a derivative with respect to the $\log k_S$. After this algebra we obtain:
\begin{equation}
\langle\zeta\zeta\rangle_B (\vec{k}_2,\vec
k_3)=\langle\zeta\zeta\rangle_0 (k_S)-\zeta^B(\vec{k}_L)
\frac{1}{k^3_S}\frac{d}{d \log k_S}[k_S^3
\langle\zeta\zeta\rangle_0(k_S)]\ ,
\end{equation}
where we have used that $\vec k_2\simeq -\vec k_3\simeq \vec k_S$ in the squeezed limit
up to sub-leading corrections. We can now multiply by $\zeta^{(B)}_{\vec k_1}$ and take
the average. The piece which is independent of the background gives no contribution, and
we are left with:
\begin{eqnarray}
\langle\zeta^{(B)}(\vec k_1)\langle\zeta\zeta\rangle_B (\vec{k}_2,\vec k_3)\rangle&=&-\langle\zeta^{(B)}(\vec k_1)\zeta^B(\vec{k}_L)\rangle
\frac{1}{k^3_S}\frac{d}{d \log k_S}[k_S^3 \langle\zeta\zeta\rangle_0(k_S)]\\
&=& -(2\pi)^3\delta^3(\vec k_1+\vec k_L) P_{k_1} P_{k_2}\frac{d \log[k^3_2 P_{k_2}]}{d\log k_2}\\
&=& -(2\pi)^3\delta^3(\vec k_1+\vec k_2+\vec k_3) P_{k_1} P_{k_2}\frac{d \log[k^3_2 P_{k_2}]}{d\log k_2}
\end{eqnarray}
where we have used eq.(\ref{two-point}).

We have thus obtained eq.(\ref{consistency}) as we wished. We stress again that {\it no}
slow roll expansion was done in this proof (which remains valid for example even if there are
sharp features in the potential), and that the only important assumption was that
there is only one clock field. This has allowed us to expand the short scale 2-point function
 in powers only
of the background field $\zeta^B$ in eq.(\ref{eq:onaback}) and not also in terms of some
other field.

We now are going to briefly introduce our effective field theory for one clock inflation,
and then to explicitly verify that the consistency relation holds with a direct calculation.

\section{Explicit Verification with the effective theory for inflation \label{sec: full verification}}
\subsection{The effective Lagrangian for single field inflation \label{sec: effective Lagrangian}}
In this section we briefly introduce the effective action for single clock
 inflation that we will use to verify the consistency relation at
 leading non trivial order in slow roll parameters. This effective action
was developed in \cite{Creminelli:2006xe, us} and we refer the reader to
those papers for a detailed explanation. The construction of the effective theory is based
on the following consideration. In a quasi de Sitter background with only one relevant
degree of freedom, there is a privileged
spatial slicing, given by the physical clock which allows us to smoothly connect to a decelerated hot 
Big Bang evolution. The slicing is usually realized by a time evolving scalar $\phi(t)$. To describe perturbations around this solution one can choose a gauge where the privileged slicing coincides with surfaces of constant $t$, {\em i.e.} $\delta\phi(\vec x,t)=0$. In this gauge there are no explicit scalar perturbations, but only metric fluctuations. As time diffeomorphisms have been fixed and are not a gauge symmetry anymore, the graviton now describes three degrees of freedom: the scalar perturbation has been eaten by the metric. One therefore can build the most generic effective action with operators
that are functions of 
the metric fluctuations and that are invariant under the linearly realized time dependent spatial
diffeomorphisms. As usual with effective field theories, this can be done in a low
energy expansion in fluctuations of the fields and derivatives.
We obtain for the matter Lagrangian \cite{Creminelli:2006xe, us}:
\begin{eqnarray}\label{Smatter_1}
S_{\rm matter} = \int  \! d^4 x  \: \sqrt{- g} &&\left[ -M^2_{\rm
Pl} \dot{H} \frac{1}{N^2} - M^2_{\rm Pl} \left(3 H^2
+\dot{H}\right)+ \right. \\ \nonumber && \left. \frac{M(t)^4}{2!}
\left(\frac{1}{N^2}-1\right)^2+\frac{c_3(t) M(t)^4}{3!}\;
\left(\frac{1}{N^2}-1\right)^3+ \right.
\\ &&\left. \nonumber +\frac{d_1(t)}{2} M^3(t) \delta N \delta E^i {}_i -\frac{d_2(t)}{2}\; M(t)^2\delta E^i {}_i {}^2
-\frac{d_3(t)}{2}\; M(t)^2 \delta E^i {}_j \delta E^j {}_i + ...
\right] \; ,
\end{eqnarray}
where we have used the ADM formalism. We have defined:
\begin{equation}
\delta N=N-1 \ ,
\end{equation}
and $\delta E_{ij}=E_{ij}-a^2 H\hat g_{ij}$ is the fluctuation in the
quantity $E_{ij}$ which is related to the
extrinsic curvature of hypersurfaces of constant $t$:
\begin{equation}
\label{extrinsic} E_{ij} \equiv N K_{ij} = \frac 12 [{\partial_t
{\hat g}}_{ij} - \hat \nabla_i N_j - \hat \nabla_j N_i] \; .
\end{equation}
Here $\hat \nabla_i$ is the derivative with respect to the spatial metric $\hat g_{ij}$.

At this point it is useful to reintroduce the full diff. invariance of the theory
by reintroducing the Goldstone boson $\pi$ of time translation with the
so called St\"u{}ckelberg trick. This amounts to performing a time diffeomorphism
in the Lagrangian of (\ref{Smatter_1}) of parameter $-\pi$, and then promoting
$\pi$ to a field which shifts under time-diffeomorphisms:
\begin{equation}
\pi\rightarrow \tilde\pi(\tilde x(x))=\pi(x)-\xi^0(\vec{x},t) \ .
\end{equation}
This procedure is explained more in detail in \cite{us}, and we refer to
it for further details.
Neglecting for the moment the terms that involve the extrinsic curvature,
we obtain:
\begin{eqnarray}\label{Smixed}
S_{\rm matter} = \int \! d^4 x  \: \sqrt{- g} &&\left[ -M^2_{\rm
Pl} \dot{H}(t+\pi)\left(\frac{1}{N^2}
\left(1+\dot\pi-N^i\partial_i\pi\right)^2-\hat
g^{ij}\partial_i\pi\partial_j\pi\right) \right. \\ \nonumber
&&\left.
- M^2_{\rm Pl} \left(3H^2(t+\pi) +\dot{H}(t+\pi)\right)+ \right.\\
&&\left.\frac{M(t+\pi)^4}{2}\left(\frac{1}{N^2}
\left(1+\dot\pi-N^i\partial_i\pi\right)^2-\hat
g^{ij}\partial_i\pi\partial_j\pi-1\right)^2 + \right. \nonumber\\
\nonumber &&\left. \frac{c_3(t+\pi)\; M(t+\pi)^4}{6}
\left(\frac{1}{N^2} \left(1+\dot\pi-N^i\partial_i\pi\right)^2-\hat
g^{ij}\partial_i\pi\partial_j\pi-1\right)^3+ ... \right] \; ,
\end{eqnarray}
Obviously, there is also the Einstein Hilbert action:
\begin{equation}
\label{EH} S_{\rm EH} = \frac 12 M_{\rm Pl}^2 \int \! d^4 x \:
\sqrt{-g} \, R = \frac 12 M_{\rm Pl}^2 \int \! d^3 x \, dt \:
\sqrt{\hat g} \, \big[ N R^{(3)} + \frac{1}{N} (E^{ij} E_{ij} -
E^i{}_i {}^2) \big] \; .
\end{equation}
No $\pi$ appears explicitly in the Einstein Hilbert action after
we perform the St\"u{}ckelberg trick because the Einstein Hilbert action
is already time diff. invariant.

The validity of this effective action is very general. It assumes only the
presence of one degree of freedom which spontaneously breaks time translation and acts as
a physical clock for the system. For example
it reproduces all known models of single field inflation. We refer
to \cite{us} for a more general discussion of this point.

We are going to verify the consistency relation with our effective Lagrangian.
Clearly, for such a complex Lagrangian, doing it in full generality is a
difficult task because of the amount of algebra that this requires.
In order to keep the complexity of the algebra at a minimum,
in the next subsections we will verify it for
some particular (and still very general) cases where each time
we neglect the contribution of some specific operators.

\subsection{Verification for the Lagrangian of the form $P(-(\partial\phi)^2,\phi)$ \label{sec: verification P(phi)}}

The Lagrangian (\ref{Smixed}) is already very general even though it omits all the operators
which in unitary gauge involve the extrinsic curvature. It in fact reproduces the Lagrangian for all the models
of inflation with a single scalar field and a Lagrangian of the form
\begin{equation}
S = \int d^4 x \sqrt{-g}\; P(-(\partial\phi)^2,\phi)\ ,
\label{eq:kinflag1}
\end{equation}
which are referred to as k-inflation \cite{Armendariz-Picon:1999rj}. This is easy to see if we write the Lagrangian in unitary gauge (which here means
$\phi(\vec x,t)=\phi_0(t)$):
\begin{equation}
S = \int d^4 x \sqrt{-g}\; P(\frac{\dot\phi_0^2}{N^2},\phi_0(t))\ ,
\label{eq:kinflag_main}
\end{equation}
which is of the form of (\ref{Smixed}). In App. \ref{app: matching theories}
we explicitly perform the matching between the parameters in
the Lagrangian  (\ref{Smixed}) and the ones in (\ref{eq:kinflag_main}).

\subsubsection{The Lagrangian at first order in slow roll
parameters}

With only one scalar degree of freedom, it is necessary to integrate out the ADM
variables $N$ and $N^i$ in order to find its
effective action. Since these variables do not have kinetic
terms, their equations of motion are algebraic. This is guaranteed
to occur because of diff. invariance. Solving for $N$ and $N^i$ in
terms of $\pi$, we can substitute these back into the Lagrangian.
This process has been
discussed extensively in \cite{Maldacena:2002vr,Chen:2006nt}, and
corresponds to removing gauge degrees of freedom.  Indeed,
counting all gauge degrees of freedom, there are exactly four
scalars in the metric and one from the matter sector (namely,
$\pi$).  We shall fix time and space diffeomorphisms by choosing
the gauge \footnote{From here on we concentrate only on scalar fluctuations. This is enough 
for verifying the consistency condition of scalar perturbations at the leading order in slow roll parameters. For higher order calculations, one should include also tensor modes.}:
\begin{equation}\label{eq: pi gauge}
\hat{g}_{ij}=a^2(t)\delta_{ij}\ .
\end{equation}
which removes two scalar degrees of freedom from the metric. Then,
we will solve for $N$ and $N^i$ removing two more, yielding one
scalar degree of freedom $\pi$ in the final action.

The equations of motion for $N$ and $N^i$ are
\begin{eqnarray}
\dd{({\cal{L}}_{\rm EH}+{\cal{L}}_{\rm matt})}{N^i}
&=& 0 \ , \\
\dd{({\cal{L}}_{\rm EH}+{\cal{L}}_{\rm matt})}{N} &=& 0 \ .
\end{eqnarray}
For the general properties of the ADM parameterization, these two
equations allow us to express $N$ and $N^i$ in terms of $\pi$. As
outlined in \cite{Maldacena:2002vr, Chen:2006nt}, we need only to
solve the constraint equations to first order in $\pi$ because we
are only interested in cubic interactions in the Lagrangian.
Solving for small fluctuations around the metric $\delta N = N -
1$ to first order in $\pi$, we find that
\begin{eqnarray}
\delta N &=&
\epsilon H \pi,\\
\partial^i N_i &=&
-\frac{\epsilon H \dot{\pi}}{c_s^2} \ ,
\end{eqnarray}
where we have defined the slow roll parameter
\begin{equation}
\epsilon=-\frac{\dot H}{H^2} \ ,
\end{equation}
and the speed of sound as:
\begin{equation}
c_s^{-2}=1-2\frac{M^4(t)}{\dot H M_{\rm Pl}^2} \; .
\end{equation}
 Plugging these back
into the Lagrangian, and concentrating on only up to the next to leading terms in slow roll parameters we obtain
\begin{eqnarray}\label{Actioncomplete}{\cal{L}} &=& {\cal{L}}_2 + {\cal{L}}_3,
\\ \label{Actioncomplete2}{\cal{L}}_2 &=& a^3 \bar M^4 \left(\dot{\pi}^2-\frac{c_s^2}{a^2}
(\partial_i\pi)^2 + 3\epsilon H^2 \pi^2 \right),\\ {\cal{L}}_3 &=& a^3
\left( C_{\dot{\pi}^3} \dot{\pi}^3 + \frac{C_{\dot{\pi}
(\partial\pi)^2}}{a^2}\dot{\pi} (\partial_i\pi)^2 + C_{\pi
\dot{\pi}^2}\pi \dot{\pi}^2 + \frac{C_{\pi (\partial\pi)^2}}{a^{2}}
\pi (\partial_i\pi)^2 + C_{\rm NL} \dot{\pi} \partial_i\pi
\partial^i\frac{1}{\partial^2}\dot{\pi} \right),
\end{eqnarray}
where for
brevity we defined $\bar M^4 \equiv \epsilon H^2 M_{\rm Pl}^2/c_s^2=2 M^4/(1-c_s^2)$. The
cubic coefficients are
\be\label{eq:full coefficients} C_{\dot{\pi}^3} &=& \bar M^4 \left(1-c_s^2\right)\left(1 +\frac{2}{3}c_3\right), \\ \nn
C_{\dot{\pi} (\partial\pi)^2} &=& \bar M^4 \left(-1 + c_s^2\right), \\ \nn
C_{\pi \dot{\pi}^2} &=& \bar M^4 H \left(-6\epsilon +\eta -2 s +3
\epsilon c_s^2
-2 \epsilon c_3 (1-c_s^2)\right),\\ \nn
C_{\pi (\partial\pi)^2}
&=& \bar M^4 H \left(\epsilon - \eta c_s^2\right), \\ \nn
C_{\rm NL} &=& \bar M^4 H \left(\frac{2 \epsilon}{c_s^2} \right)\;
, \ee where we have defined the other slow roll parameters as:
\begin{eqnarray}
&&\eta=\frac{\dot\epsilon}{\epsilon H} \\ \nn \nonumber &&
s=\frac{\dot{c_s}}{c_s H} \ .
\end{eqnarray}
Notice that the final term is a non-local interaction term which
arises from the fact that we have written a gauge fixed Lagrangian. Obviously, all gauge invariant observables will be
local. At next to leading order in slow roll parameters we have five distinct cubic
operators which give rise to the five distinct shapes for the
three point function.

\subsubsection{2-point function and its tilt}

In order to verify the consistency relation at first order in slow
roll parameters, we need to compute the 2-point function
(\ref{two-point}) for $\zeta$ at first order in slow roll. In
order to do this, we need to find the relationship between $\pi$
and $\zeta$. This is given in App. \ref{app: non linear zeta_pi}
at the non linear level. Here we just need the relationship at
linear level. It is rather straightforward to realize that in
Fourier space:
\begin{equation}
\zeta_k=-H_*\pi_k \; ,\ {\rm at\; linear\; level} \; .
\end{equation}
where the $_*$ means that the quantity has to be evaluated at
horizon crossing, that is when $\omega^2=k^2/a(t_*)^2=H(t_*)^2$.
This relationship can be understood as follows. At linear level no
space derivative can appear. Then, the relationship can be derived
thinking of very long wavelengths. As we mentioned in sec.
\ref{sec: formal proof}, on large scales $\zeta$ corresponds to
the relative expansion between separate unperturbed FRW universes.
Similarly, on large scales, $\pi$ represents the time-delay
between the same separate unperturbed FRW universes. Therefore, we
deduce that $\zeta=-H_* \pi$, at least at linear level, as is
verified in App. \ref{app: non linear zeta_pi}. The sign depends
on our definition of $\pi$.

Now, we would need to find the wavefunction for the mode
$\pi^{cl}_k(t)$ defined by the relationship $\pi_k = \pi^{cl}_k
\hat{a}_k + \pi^{cl}_{-k} {}^* \hat{a}^\dag_{-k}$. Doing this at first
order in slow roll parameters is a rather tedious task, that we
perform in App. \ref{app: waveeq}. However, one can easily find
the solution in exact de Sitter from the quadratic Lagrangian
(\ref{Actioncomplete2}):
\begin{eqnarray}\label{eq:modenorm}
\pi^{cl}_k(\tau) &=& i \frac{1}{2 \sqrt{\epsilon k^3 c_s } M_{\rm Pl}}
   (1+ i k c_s \tau)e^{-i k c_s \tau} \ , \\ \nonumber
\zeta^{cl}_k(\tau)&=&-H\pi^{cl}_k \ ,
\end{eqnarray}
where we have imposed the usual Minkowski vacuum at early
times. Here $\tau$ is the conformal time. The power spectrum
becomes:
\begin{equation}
P_k=\frac{H_*^2}{4\; \epsilon\; c_{s,*}\; M^2_{\rm Pl}}\frac{1}{k^3} \
.
\end{equation}
Here this quantity is evaluated at $t_*$ in order to minimize the
error we introduced in evaluating the wavefunction in de Sitter. The
dependence on $t_*$ induces an additional momentum dependence. It
is convenient to parameterize it by saying that the total
correlation function has the form $k^{-3+n_s}$, where
\begin{eqnarray} \label{tilt}
n_s&=&\frac{d\;\log k^3 P_k}{d\;\log k}=k\frac{d}{d\;k}\log\left(\frac{H^4_{*}}{\dot{H}_*
c_{s*}M_{\rm Pl}^2}\right)\sim\frac{1}{H_*}\frac{d}{d\;
t_*}\log\left(\frac{H^4_{*}}{\dot{H}_*
c_{s*}M^2_{\rm Pl}}\right)\nonumber\\
&=&4\frac{\dot{H_*}}{H^2_*}-\frac{\ddot{H}_*}{\dot{H}_*
H_* }-\frac{\dot{c}_{s*}}{c_{s*}H_{*}}=-2\epsilon-\eta-s \ .
\end{eqnarray}

In App. \ref{app: waveeq} we explicitly verify that this is the
correct result.
Notice also that, as anticipated, after horizon crossing the commutator
$[\zeta,\dot\zeta]\rightarrow 0$ exponentially fast in cosmic time.

\subsubsection{3-point function at leading order in slow roll parameters \label{sec:explicitcalc_leading}}

At leading order in slow-roll, remarkably little work or subtlety
is involved, since mixing with gravity, time-dependence of
coefficients, corrections to pure de Sitter wave functions, and
corrections to the evolution of comoving time $\tau$ can all be
ignored. At this order we have a contribution only from two
operators: $\dot\pi(\partial\pi)^2$ and $\dot\pi^3$. The calculation
of non-Gaussianities in the power spectrum of fields such as $\pi$
from their interactions in the Lagrangian is a standard
\cite{Maldacena:2002vr} calculation of the expectation value
$\langle \pi^3 \rangle$.
\begin{equation}
\langle \pi^3(t_0)  \rangle =- i \int_{-\infty}^{t_0} dt \langle [
\pi^3(t_0), {\mathcal{H}}_{\textrm{int}}(t) ] \rangle
\label{eq:vev}
\end{equation}
where $t_0$ is some time well after all modes have exited the
horizon. At leading order in interaction we have ${\cal{H}}_{\rm
int}=-{\cal{L}}_{\rm int }$. We will focus first on the
interaction $\mathcal{L}_{\textrm{int}} = -2 M^4 \int d^3x a^3
\dot{\pi} (\partial_i \pi/a)^2$.  Equation (\ref{eq:vev}) is
evaluated for the explicit $\pi$ operators in terms of the
interaction picture wavefunctions, $\pi_k = \pi^{cl}_k \hat{a}_k +
\pi^{cl}_{-k} {}^* \hat{a}^\dag_{-k}$.
\begin{eqnarray}
 \langle \pi_{k_1} \pi_{k_2} \pi_{k_3} \rangle &=&
  i\; C_{\dot{\pi}\partial \pi^2} (2\pi)^3
\delta^3\left( \sum_i \vec{k}_i\right) \pi^{cl}_{k_1}(0)
\pi^{cl}_{k_2}(0) \pi^{cl}_{k_3}(0) \nonumber \\
   & & \cdot \int_{-\infty}^{0}
   \frac{d\tau}{H \tau}  \frac{d}{d\tau}\pi^{cl}_{k_1} {}^* (\tau)
\pi^{cl}_{k_2} {}^* (\tau)  \pi^{cl}_{k_3} {}^*
(\tau)
 \left( \vec{k}_2 \cdot \vec{k}_3 \right) + \mathrm{permutations} + \mathrm{c.c.} \nonumber\\
&&
\end{eqnarray}
where the sum above includes all symmetric permutations of the
three momenta. Inserting the expression for the wavefunction in
eq. (\ref{eq:modenorm}) into the above expression for the
three-point function, we find
\begin{eqnarray}
\langle \pi_{k_1} \pi_{k_2} \pi_{k_3} \rangle &=&
   -i M^4 (2\pi)^3 \delta^3(\sum_i \vec{k}_i)
      \frac{1}{32 c_s^3 \epsilon^3 H M^6_{\rm Pl}\prod_i k_i^3} \nonumber \\
  &\times&
  \int_{-\infty}^0 d\tau c_s^2k_1^2 (k_2 \cdot k_3)
     (1-i k_2 c_s \tau) ( 1 - i k_3 c_s \tau) e^{i \sum_i k_i c_s \tau}
  \nonumber \\
  &+& \mathrm{permutations} + c.c.
  \label{eq:xxt}
\end{eqnarray}
The above integral converges if we rotate the path of integration
slightly into the complex plane to pick out the interacting vacuum
state in the infinite past, so that $\tau \rightarrow -\infty(1-i
\epsilon)$. We can use the constraint $\sum \vec{k}_i=0$ to
simplify the above expression further.  Since $\sum \vec{k}_i=0$,
it follows that $k_1 \cdot k_2 = \half (k_3^2 - k_1^2 - k_2^2)$,
and permutations thereof.  This makes it easy to eliminate the
inner products in favor of the magnitudes of the $k_i$'s. All the
cubic operators contribute to the three-point function via similar
computations to the integral above. It is useful to define a basis
of symmetric products of $k_i$ wavevectors.  We choose the
combinations
\begin{eqnarray}
K_1 &=& k_1 + k_2 + k_3 \\
K_2 &=& \left( k_1 k_2 + k_2 k_3 + k_3 k_1 \right)^{1/2} \\
K_3 &=& \left( k_1 k_2 k_3 \right)^{1/3}
\end{eqnarray}
and any symmetric polynomial in $k_i$'s can be decomposed in terms
of these.
Equation (\ref{eq:xxt}) is now easy to evaluate and express in
terms
 of the symmetric variables $K_i$.
In general, the contribution from an operator $\mathcal{O}$ in the Lagrangian which is cubic
in $\pi$'s can be put in the form
\begin{eqnarray}\label{eq:generalnongauss}
\langle\pi_{k_1}\pi_{k_2}\pi_{k_3}\rangle&&=i \int d^3 x \int_{-\infty}^0 a d\tau  \sqrt{-g}\; \langle \left[
\pi_{k_1} (\tau) \pi_{k_2} (\tau) \pi_{k_3} (\tau), C_{\mathcal{O}}\mathcal{O}
\right] \rangle
  \equiv \\ \nonumber &&\equiv C_{\mathcal{O}}
  \left|\prod_{i=1}^3 \pi^c_{k_i}(0)\right|^2
 (2\pi)^3 \delta^3(\sum_i \vec{k}_i)
 \frac{A_{\mathcal{O}}(c_s, H,K_1,K_2,K_3)}
{K_1^3} \nn
\end{eqnarray}
where in this case the $C_{\mathcal{O}}$ are $C_{\dot{\pi}^3}$ and
$C_{\dot{\pi} \partial \pi^2}$ evaluated at zeroth order in slow
roll parameters . Regardless of the size of $C_{\mathcal{O}}$, it
is straightforward to calculate the leading-order contribution to
any shape $A_{\mathcal{O}}$ by using the computational procedure
described above.  For $\mathcal{O}$ = $\dot{\pi}^3$ and $\dot{\pi}
(\partial_i\pi)^2$, they are
\begin{eqnarray}
 \label{eq:shapes}
 A_{\dot{\pi} (\partial \pi)^2}&=&\left(\frac{c_s}{H}\right) \left(
24 K_3^6- 8 K_2^2 K_3^3 K_1- 8 K_2^4 K_1^2+22 K_3^3 K_1^3- 6 K_2^2 K_1^4+ 2 K_1^6\right)\\ \nn
 A_{\dot{\pi}^3}&=&\left(\frac{c^3_s}{H}\right)\left(24
  K_3^6\right)
\end{eqnarray}
 To leading order in slow roll parameters, then, the $\pi$
3-point function is
\begin{equation}
\langle \pi_{k_1} \pi_{k_2} \pi_{k_3} \rangle =
   \left|\prod_{i=1}^3 \pi^{cl}_{k_i}(0) \right|^2
(2\pi)^3 \delta^3( \sum_i \vec{k}_i)
   \frac{ 2 M^4 \left(-A_{\dot{\pi} \partial \pi^2} + ( 1 + 2 c_3/3 )
   A_{\dot{\pi}^3}\right)}{K_1^3}
\end{equation}
To find the 3-point function for $\zeta$ ($\zeta$ is constant
outside of the horizon, and therefore it is the relevant quantity
for observation), we need the relation between $\pi$ and $\zeta$.
This can be found performing the diffeomorphism that connects the
$\zeta$-gauge (\ref{zetagauge}) and the $\pi$-gauge (\ref{eq: pi
gauge}). This is given in App. \ref{app: non linear
zeta_pi}~\cite{Maldacena:2002vr}:
\begin{eqnarray}
\zeta &=& -H \pi +H \pi \dot{\pi}
  + \half \dot{H} \pi^2 + \alpha   \\
4\alpha &=&\frac{1}{a^2} \left(- \partial_i \pi \partial_i \pi +
  \partial^{-2} \partial_i \partial_j (\partial_i \pi \partial_j
  \pi)\right)
\end{eqnarray}
The $\alpha$ term is proportional to spatial derivatives of $\pi$,
so it vanishes outside the horizon and will not contribute in the
expression for the three-point function. The $\half \dot{H} \pi^2$ in
$\zeta$ contributes to the $\zeta$ three-point function through
$\langle \zeta^3 \rangle \supset \langle H^2 \dot{H} \pi^4
\rangle$, which is sub-leading in slow roll expansion with respect
to the terms we are keeping here. Finally, the $H \pi \dot{\pi}$
in $\zeta$ contributes through $\langle \zeta^3 \rangle \supset
\langle H^3 \pi^3 \dot{\pi} \rangle$, but this also vanishes at
this order in slow-roll since in exact de Sitter space $\dot{\pi}$
vanishes outside the horizon. The full three-point function for
$\zeta$ to leading order in $\epsilon$ is therefore
\begin{eqnarray}
\langle \zeta_{k_1} \zeta_{k_2} \zeta_{k_3} \rangle &=&
  - H^3 \langle \pi_{k_1} \pi_{k_2} \pi_{k_3} \rangle  \nn
  =-
\pi^3 \delta^3 ( \sum_i \vec{k}_i)
  \frac{H^3M^4 \left(- A_{\dot{\pi} \partial \pi^2}
  + (1 + 2 c_3 /3) A_{\dot{\pi}^3} \right) }{4 c_s^3
 M_{\rm Pl}^6 \epsilon^3 K_3^9 K_1^3} \\
  &=&\pi^3 \delta^3( \sum_i \vec{k}_i )
  H^3\dot{H}(1-c_s^2) \frac{\left(- A_{\dot{\pi} \partial \pi^2}
  + (1 + 2 c_3 /3) A_{\dot{\pi}^3} \right) }{8 c_s^5
 M_{\rm Pl}^4 \epsilon^3 K_3^9 K_1^3}
\label{threepointsleading}
\end{eqnarray}
where in the last step we have used $M^4=\dot H M_{\rm
Pl}^2(1-1/c_s^2)/2$. In App. \ref{app: zeta cosntant} we show
using our generic effective theory (\ref{Smatter_1}) that $\zeta$
is constant outside of the horizon at fully non linear level.

\subsubsection{3-point function at next to leading order in slow roll
parameters \label{sec:explicitcalc_next to leading} } At next to
leading order in slow-roll,
 the computation of the non-Gaussianities involves considerably
more work, and much of the computational advantage of using the
$\pi$ field instead of the $\zeta$ field is lost. This is
understandable, as the Goldstone boson is advantageous for
describing the UV physics, while here we are also considering some
IR effects. At any order in slow roll, the contribution from an
operator $\mathcal{O}$ is still exactly given by the general
expression (\ref{eq:generalnongauss}).  The calculation of
$C_{\mathcal{O}}$, $\pi^{cl}_k(0)$ and $A_{\mathcal{O}}/K_1^3$
should be carried out at the required order, but they are all
independent of each other.

We have just calculated the coefficients $C_{\mathcal{O}}$ at the
next leading order in slow roll parameters. 
The next to leading order corrections involve three new operators
that were not present at leading order. For these three operators,
the contribution to the non-Gaussianities is actually quite easy.
Their coefficients are already suppressed by slow-roll, so we can
simply use the leading order contributions to $A_{\mathcal{O}}$, and be done. The operators
$\dot{\pi}^3$ and $\dot{\pi} (\partial \pi)^2$ that were present at
leading order are more complicated.   In addition to the slow-roll
corrections to $C_{\dot{\pi}^3}$ and $C_{\dot{\pi} (\partial
\pi)^2}$, there are several corrections inside the integral of
equation (\ref{eq:generalnongauss}). Such corrections have been
treated before\cite{Chen:2006nt, Stewart:1993bc}, but we will not need them in our calculation.
We can write the integral to evaluate symbolically as
\begin{eqnarray}
\langle \pi_{k_1} (0) \pi_{k_2} (0) \pi_{k_3}(0) \rangle &=&
  i \int d^3 x \int_{-\infty}^0 \textrm{[ MEASURE ] [ TIME-DEPENDENCE ]} \nn\\
 &\times& \textrm{ [ DERIVATIVES ] [ MODES ] + c.c. + symm. } \ .
\end{eqnarray}
It is not very instructive to write the full result, because we
cannot put it in closed form in any case. Instead, we turn
directly to verify the consistency relation, as we can do it
without many complication. In fact, there is a nice trick that we can use
in the squeezed limit to avoid these complications.  To understand why this
is the case, let us first ask which operators contribute in the squeezed
limit if we work in the $\zeta$ gauge instead of the $\pi$ gauge.
A key insight is that $\zeta$ does not evolve outsize the horizon,
and so operators like $\dot{\zeta}^3$ and $\dot{\zeta} (\partial
\zeta)^2$, with a derivative on each $\zeta$, do not contribute in
the squeezed limit.
If we imagine doing the calculation directly for $\zeta$, we see
that the contribution from the operator $\dot\zeta^3$ will be
proportional to the derivative of $\zeta_{k_L}$, and therefore
will be negligible in the squeezed limit. However, the same reasoning
does not apply to $\pi=-\zeta/H+...$, which
evolves outside the horizon, and therefore the operators
$\dot{\pi}^3$ and $\dot{\pi} (\partial \pi)^2$ still contribute in
the squeezed limit. However, the interaction picture wavefunction,
which follows the linear equation of motion, is \emph{almost}
constant outside the horizon. In particular, using that
$\zeta^{cl}=-H\pi^{cl}$, we can write:
\begin{equation}
\dot{\pi}^{cl}  = -\frac{\dot{\zeta}^{cl}}{H} + H \epsilon
\pi^{cl}
\end{equation}
We emphasize that this relation is true for the classical
solutions, but not for the $\pi$ and $\zeta$ operators in general,
and that one should only use the above substitution under the
integral in the computation of the 3-point functions
(\ref{eq:generalnongauss}). But this is enough for us. We have to
compute corrections only to terms that were already present at
leading order in slow roll, $\dot{\pi}^3$ and $\dot{\pi} (\partial
\pi)^2$. Focussing on the latter of these for the moment, we make
the replacement
\begin{equation} \label{ZetaSubstition}
\dot{\pi}^{cl} (\partial \pi^{cl})^2 \rightarrow -\frac{1}{H}
\dot\zeta^{cl} (\partial \pi^{cl})^2
  + H \epsilon \pi^{cl} (\partial \pi^{cl})^2
\end{equation}
The second term on the RHS is now suppressed by $\epsilon$, so its
shape can be calculated without any slow-roll corrections, as we
shall soon do. The first term is clearly going to be sub-leading in
the squeezed triangle limit, because there is one derivative on each
field, and therefore we expect a suppression
of order $k_L/k_S$ with respect to the leading behavior. However,
since, in the actual computation, this term is under an integral
over $\tau$, it might not be obvious to the reader that this
contribution really vanishes.   This last fact can be explicitly
verified, recovering the $k$ dependence of the contribution of the
first term on the RHS of (\ref{ZetaSubstition})
in the squeezed limit. Let us see how we can
do this. We have to evaluate the following integral:
\begin{eqnarray}\label{zetaint}
 \langle \pi_{k_1} \pi_{k_2} \pi_{k_3} \rangle &=&
  -i\; C_{\dot{\pi}\partial \pi^2} (2\pi)^3
\delta^3( \sum_i \vec{k}_i) \pi^{cl}_{k_1}(0)
\pi^{cl}_{k_2}(0) \pi^{cl}_{k_3}(0) \nonumber \\
   & & \cdot \int_{-\infty}^{0}
   \frac{d\tau}{H^2 \tau} \frac{d}{d\tau}\zeta^{cl}_{k_1} {^*} (\tau)\
\pi^{cl}_{k_2} {}^* (\tau)  \pi^{cl}_{k_3} {}^* (\tau)
 \left( \vec{k}_2 \cdot \vec{k}_3 \right) \ + \ {\rm c.c.}\ + \ {\rm symm.}\ \ .\nonumber\\  &&
\end{eqnarray}
where we have contracted $\pi_{k_1}$ with $\dot\zeta_{k}$ as an illustrative
example.
We can divide the integral over conformal time into two regions:
``early times'', when $k_S \gg k_L \gtrsim  a H$ and the physical
wavelengths are inside the horizon; ``late times'', when $k_L \ll
a H$ and the long physical wavelength is outside the horizon.
Now, the contribution from early times is negligible due to the
rapid oscillations from the exponential factor $e^{i k_S c_s
\tau}$ \footnote{In practice, the oscillatory damping is even more
obvious since one analytically continues into the complex plane
where it becomes exponential damping.}. In the remainder of the
integral, $k_L$ can safely be taken as small as we like. In this
limit, then the Hankel function of the field which has the long
wavelength mode can be expanded in the small argument limit. It is
clear that this expansion will bring powers of $k_L$ in the
numerator, making the contribution of this term negligible in the
squeezed limit\footnote{This is the step of the proof that fails
for the term $\dot{\pi} (\partial \pi)^2$ but holds for $\dot{\zeta}
(\partial \pi)^2$.}.
But let us explicitly see how this works considering the particular
case where $k_1=k_L$, and $k_2\simeq k_3=k_S$. In this case we can
expand the wave function of $\zeta$, as done in App. \ref{sec:long
wave limit}. Considering only the parametric dependence on
momenta, the first term from equation (\ref{zetaint}) contributes
as
\begin{eqnarray}
 \langle \pi_{k_L} \pi_{k_s} \pi_{k_s} \rangle &\propto&
  \pi_{k_L}^{cl}(0)
\pi_{k_s}^{cl}(0) \pi_{k_s}^{cl}(0)
   \cdot \int_{-\infty}^{0}
   \frac{d\tau}{\tau} \frac{1}{\tau}\tau^2 k_L^{1/2} k_S^2
\pi^{cl*}_{k_s} (\tau) \pi^{cl*}_{k_s} (\tau) \nonumber\\
&\propto& \frac{1}{k_L^{3/2}}\frac{1}{k_S^{3}}
   \cdot \int_{-\infty}^{0}
   \frac{d\tau}{ \tau} \frac{1}{\tau}\tau^2 k_L^{1/2}
   \left(k_S (-\tau)^{3/2}H_\nu^{(1)}(-c_s k_S \tau (1+s))\right)^2
   \nonumber \\
   &\propto&
   \frac{1}{k_L^3}\frac{1}{k_S^3}\left(\frac{k_L}{k_S}\right)^2
   \int_{-\infty}^0 dy \ y^2 (H_\nu^{(1)}(-y))^2 \ .
\end{eqnarray}
where in the first passage we have used the long wavelength limit
of $\dot\zeta$ given in App. \ref{sec:long wave limit}, in the
second we have used the explicit form for the wave functions at
first order in slow roll parameters given in (\ref{piwavefinc}),
and in the third we have changed variable of integration to $y=c_s
k_S \tau(1+s)$. We see that the remaining integral is just a
numerical factor, and the $k$ dependence of the contribution is
suppressed with respect to the leading one in the consistency
relation by a factor of $(k_L/k_S)^2$. It is clear that the same
arguments apply to the other terms of eq.(\ref{zetaint}). We avoid
showing the very similar argument for the operator
$\dot\pi^3$ except there is an extra factor of 3, since
$\pi_{k_L}$ can be put on any of the three $\dot{\pi}$'s. Notice
that this works out so that the terms in $c_3$ coming from
$(1/N^2-1)^3$ do not contribute in the squeezed limit. This is
very important for the consistency relation to be true, as no
term from $(1/N^2-1)^3$ contributes to the two point function, and therefore
to its tilt.


Applying this argument, we find that the $\pi$ three-point
function in the squeezed limit ($k_1\rightarrow 0$) at next to leading order is
\begin{eqnarray}
\langle \pi_{k_1} \pi_{k_2} \pi_{k_3} \rangle &=&
  \left|\prod_{i=1}^3 \pi^{cl}_{k_i}(0)\right|^2 (2\pi)^3
   \delta^3( \sum_i \vec{k}_i ) \nn\\
&\times& \frac{ (C_{\dot{\pi}^3} 3 H \epsilon + C_{\pi
\dot{\pi}^2} )
 A_{\pi \dot{\pi}^2}
+ (C_{\dot{\pi} (\partial \pi)^2} H \epsilon + C_{\pi (\partial \pi)^2} )
  A_{\pi \partial \pi^2} }{K_1^3}
\end{eqnarray}
The shapes $A_{\pi (\dot{\pi})^2}$ and $A_{\pi (\partial \pi)^2}$
can be calculated using the method described in section
\ref{sec:explicitcalc_leading}. We obtain
\begin{eqnarray} \label{eq:2results}
      A_{\pi (\partial \pi)^2}&=&\left( \frac{c_s}{H^2} \right)\left(4K_2^2 K_3^3 K_1+ 4K_2^4 K_1^2- 2K_3^3 K_1^3- 6 K_2^2 K_1^4 + 2 K_1^6\right) \\ \nn
  A_{\pi \dot{\pi}^2}&=&\left(\frac{c_s^3}{H^2} \right)\left(4K_2^2 K_3^3 K_1+ 4K_2^4 K_1^2- 8 K_3^3 K_1^3\right)
\end{eqnarray}
At next to leading order, $\zeta = -H \pi + H \pi \dot{\pi} +
\half \dot{H} \pi^2 + 4\alpha$. Then, in Fourier space, $\zeta$ is
of the form $\zeta_k = -H \pi_k + H (\pi * \dot{\pi})_k + \half
\dot{H}  (\pi*\pi)_k$, where $(\pi*\pi)_k = \int d^3 q \; \pi_{k-q}
\pi_q / (2 \pi)^3$ is a convolution. As before, $\dot{\pi}$ outside the
horizon can be replaced by $H \epsilon \pi$ and the terms contained
in $\alpha$ are irrelevant, so outside the horizon we can take $\zeta_k = - H
\pi_k - \half \dot{H} (\pi
* \pi)_k$. Considering the contribution from the field
redefinition, the full three-point function
for $\zeta$ in the limit $k_1=k_L\rightarrow 0$ is
\begin{eqnarray}\label{partial 3-point} \nn
\langle \zeta_{k_1} \zeta_{k_2} \zeta_{k_3} \rangle &=&
   -H^3 \langle \pi_{k_1} \pi_{k_2} \pi_{k_3} \rangle
   - \frac{\dot{H}}{H^2} (2\pi)^3 \delta^3( \sum_i \vec{k}_i)
     \left( P_{k_1} P_{k_2} + P_{k_2} P_{k_3} + P_{k_1} P_{k_3}\right) \\
   &\simeq& -(2\pi)^3 \delta^3( \sum_i \vec{k}_i)
  P_{k_S} P_{k_L}
\\ && \times
  \left( \left|\pi^{cl}_{k_S}(0) \right|^2
 \frac{ (C_{\dot{\pi}^3} 3 H \epsilon + C_{\pi \dot{\pi}^2} )
 A_{\pi \dot{\pi}^2}
+ (C_{\dot{\pi} (\partial \pi)^2} H \epsilon + C_{\pi (\partial \pi)^2} )
  A_{\pi (\partial \pi)^2} }{8 k_S^3 H}
  - 2 \epsilon \right) \nonumber
\end{eqnarray}
Here $P_k$ is the power spectrum defined in eq.(\ref{two-point}).
In the first line, the second term in the right hand side comes from the
field redefinition. In the second line we have taken the squeezed limit,
and used
that
$P_{k_S}\ll P_{k_L}$.

It is straightforward to check that the last term above in
parentheses is equal at next to leading order to $n_s$. In the
squeezed limit we have,
\begin{equation}
A_{\pi (\partial \pi)^2} = \frac{48 c_s k_S^6}{H^2}\;,
\ \ \ \ A_{\pi \dot{\pi}^2} = \frac{ 16 c_s^3 k_S^6}{H^2}\ .
\end{equation}
From equation (\ref{eq:modenorm}),
\begin{equation}
|\pi^{cl}_{k_S}(0)|^2 =
\frac{1}{4 \epsilon k_S^3 c_s M_{\rm Pl}^2}\; ,
\end{equation}
and from equation
(\ref{eq:full coefficients}) we have
\begin{equation}
C_{\dot{\pi}^3} 3 H
\epsilon + C_{\pi \dot{\pi}^2} = M_{\rm Pl}^2 H^3 \epsilon( - 2 s
- 3 \epsilon + \eta)/c_s^2\;, \ \ \  \  C_{\dot{\pi} (\partial \pi)^2} H
\epsilon + C_{\pi (\partial \pi)^2} = M_{\rm Pl}^2 H^3 \epsilon
(\epsilon - \eta)\;.
\end{equation}
Substituting these expressions in (\ref{partial 3-point}), we find
\begin{eqnarray}
\langle \zeta_{k_1} \zeta_{k_2} \zeta_{k_3} \rangle
&\simeq& -(2\pi)^3 \delta^3( \sum_i \vec{k}_i)
  P_{k_S} P_{k_L} (-2\epsilon -s -\eta)= -(2\pi)^3 \delta^3( \sum_i \vec{k}_i)
  P_{k_S} P_{k_L} n_s
\end{eqnarray}
which is exactly of the form of eq.(\ref{consistency}) if we
identify $k_1=k_L\rightarrow 0$ and $k_2\simeq k_3=k_S$
\footnote{The 3-point function produced by a scalar field with Lagrangian of the form
$P(-(\partial\phi)^2,\phi)$ had already
been studied in \cite{Chen:2006nt}, where it was originally found that the consistency
relation was violated by terms of the form $\epsilon/c_s^2$.  However, after
the publication of this paper, the authors of \cite{Chen:2006nt} revised their calculations
and found some algebra mistakes.  The new results of \cite{Chen:2006nt} agree with ours
completely.}
.
This
ends the verification of the consistency relation at first order
in slow roll parameters for the particular case we have considered
in this section.

\subsection{Verification for the Operator $\delta N \delta E^i{}_i$}

We now turn to the verification of the consistency relation in the case where the
operators which involve the extrinsic curvature are important. Since
these terms are higher derivative, in general they do not give rise to
the leading contribution. However this is not necessarily the case if
we are near de Sitter. As we explain more in detail in \cite{us}, the
leading gradient term to the $\pi$ field is fixed by the symmetries of
FRW to be equal to $M_{\rm Pl}^2\dot H (\partial_i\pi)^2$. This term goes
to zero as $\dot H$ vanishes, and this allows higher derivative
terms to become the leading ones. This is the case of the terms
in $\delta E^i {}_i^2$ and $\delta E^i {}_j \delta E^j {}_i$, which, upon
reinsertion of the $\pi$, give rise to:
\begin{equation}
\label{di_pi}  -\frac12
d_2 M^2 \, \delta E^i {}_i {}^2 - \frac12 d_3 \, M^2
\, \delta E^{ij} \delta E_{ij}  \; \rightarrow   - \frac12 \bar d
M^2 \, \frac{1}{a^4}(\di_i^2 \pi)^2
    \; ,
\end{equation}
where $\bar d=d_2+d_3$.
The situation
for the operator $\delta N \delta E^i {}_i$ is slightly different. Upon
reintroduction of the $\pi$, this term gives a non trivial contribution to
the action only either through its mixing with gravity, or because of the
presence of the Hubble constant. Neglecting metric fluctuations and
concentrating only on the $\pi$ terms, we obtain:
\begin{equation}
-\frac{d_1}{2} M^3 \dot \pi \frac{1}{a^2} \partial^2 \pi \to
\frac{d_1}{2} M^3 \frac{1}{2 a^2} \frac{d}{dt}(\di_i \pi)^2
 \to -\frac{H d_1 M^3}{4} \frac{1}{a^2} (\di_i \pi)^2 \; ,
\end{equation}
where we have performed an integration by parts. Since this two derivative
gradient term is suppressed by $H$ it is in general negligible unless
$\dot H$ is very small. This is the reason why, in the former section,
we completely neglected these terms. However, even though by a small amount, these operator do
contribute to the 2-point function and to its tilt, and therefore, if the consistency relation
is true, they must give a contribution
to the 3-point function in the squeezed limit. This in principle should be checked
\footnote{At this point one could wonder of the particular regime in which we are close to de Sitter and in which the coefficients of the extrinsic curvature operators we consider are tuned to be zero. Then other higher derivative terms can become important.
However, as we show in \cite{us}, when this is the case, the theory becomes strongly coupled in
the infrared, and our effective field theory description breaks down. We therefore conclude that
it is enough to restrict ourself to study the operators we consider.}.
As it should be clear from the study of the former
section, the full study of the three point function in the squeezed limit in
the case we include all these operators is clearly very long and tedious.
Furthermore, it is true that these new operators become important only if we are
close to de Sitter. For these reasons, in the rest of this section
we decide to make some simplifying assumptions. First of all, we restrict
to de Sitter background, so that we make the importance of these operators as large as possible.
Second, for each operator,
we decide to restrict ourself at verifying the consistency relation for the minimal
case which is still non-trivial, setting all the unnecessary operators to
zero. In summary, we consider two separate cases: one where we set to zero the operators $\delta E^i {}_i {}^2$ and $\delta E^i {}_j \delta E^j {}_i$, and we keep the operators $(1/N^2-1)^2$
and $\delta N \delta E^i {}_i$; and the other where we set to zero  $\delta N \delta E^i {}_i$ and  $\delta E^i {}_i {}^2$ and
we keep $(1/N^2-1)^2$ and $\delta E^i {}_j \delta E^j {}_i$
\footnote{The case where we keep $(1/N^2-1)^2$ and $\delta E^i {}_i {}^2$ and set the other operators to zero would be very similar to the second case we consider, and therefore we
avoid to explicitly perform the calculation for this case.}. We then start by considering
the following action:
\begin{equation}
S_{\rm matter}= \int d^4x\; \sqrt{-g} \left[\frac{M^4(t)}{2} \left(\frac{1}{N^2}-1\right)^2  + \frac{d_1(t)}{2} M^3(t)
 \delta N \delta E_i^i\right]\; . \label{eq: action delta N}
\end{equation}
Notice that we have set to zero the operator $(1/N^2-1)^3$
because it is not strictly necessary for
a non trivial check of the consistency relation. In the squeezed limit, the consistency relation
involves the deviation from scale invariance of the two point function. Even though we are
in de Sitter, still we obtain a scale dependence from
the time dependence of the coefficients in (\ref{eq: action delta N}), which allows us to have
a non trivial check. To this purpose, we keep only the first time derivative and drop
higher ones.

As we did in the former section, we need to find the effective action for the relevant scalar
degree of freedom. This involves reinserting the $\pi$, writing down the constraint equations,
gauge fixing and solving them in terms of $\delta N$ and $N_i$, and finally plugging back into
the gauge fixed action. This part proceeds as in the former section, with just the algebra being
a bit more complicated.

Reinsertion of the $\pi$ in the operators $\delta E^i {}_i$ leads to (see App.  \ref{App: deltaNdeltaE} for a detailed derivation):
\begin{eqnarray}
\delta E_i^i &\rightarrow& \delta E_i^i - 3 H \dot{\pi}
     + 3 H \dot{\pi}^2 - \half \d_t ( 2 \d_i \pi N^i + (\d \pi)^2)
  \nn\\
  &&- \d^2 \pi - \d_i (2 \delta N \d^i \pi
  + 3 N^i \dot{\pi} - 2 \d^i \pi \dot{\pi}) \, .
\end{eqnarray}
The constrained equation can be solved in the spatially flat gauge we use (eq.~(\ref{eq: pi gauge})) to give:
\begin{eqnarray}
\delta N &=& \frac{d_1(t) M^3(t)}{ d_1(t)M^3(t)+4HM_{\rm Pl}^2}\; \dot\pi
\end{eqnarray}
\begin{eqnarray}
\partial_i N^i
  &=&
-\frac{24 d_1(t) H^2 M_{\rm Pl}^2 M(t)^3+32 H M_{\rm Pl}^2 M(t)^4 +3 d_1(t)^2 H M(t)^6 }
{\left(4 H M_{\rm Pl}^2 +
   d_1(t) M^3(t)\right)^2}\;\dot\pi  \nn\\
  &&- \frac{d_1(t) M^3(t)}{4H M_{\rm Pl}^2 + d_1(t) M^3(t)}\;a^{-2} \d^2 \pi
\end{eqnarray}
Plugging back in (\ref{eq: action delta N}) we obtain the final action.

\subsubsection{2-point function and its tilt}
At quadratic level, the action is:
\be
S_2 &=& \frac{1}{2}\int d^3x d\tau f^2 ( \pi'^2 - c_s^2  (\partial_i \pi)^2 )
\en
where we have directly passed to conformal time. $f$ and $c_s$ are defined by:
\begin{eqnarray}\label{eq:f and c_s}
f^2 \tau^2 &=&  M_{\rm Pl}^2
\frac{64 M^4 M_{\rm Pl}^2 + 6 d_1 M^3 (8 H M_{\rm Pl}^2 + d_1 M^3)}{ (4H M_{\rm Pl}^2 +
     d_1 M^3)^2}\ , \\ 
c_s^2 &=&-
   \frac{ d_1 M^3 (4H M_{\rm Pl}^2 + d_1 M^3)+4 M_{\rm Pl}^2 \d_t(d_1 M^3)}{\left(32 M^4 M_{\rm Pl}^2 + 3 d_1 M^3
   (8 H M_{\rm Pl}^2  + d_1 M^3)\right)} \; ,
\end{eqnarray}
where for brevity we have stopped explicitly showing that $M$ and $d_1$ depend on time. The quadratic
action is the one of a scalar field with a speed of sound different from one. Using the result of
the former section, we can quickly compute the power spectrum and its tilt:
\begin{equation}
P_k=\frac{H^2}{2 f^2 \tau^2 c_s^3}\frac{1}{k^3}\, ,
\end{equation}
\begin{eqnarray}\label{eq:tilt2}
n_s \simeq \frac{d \log k^3 P_k}{d \log k}&=&
\frac{16 M^2_{\rm Pl} \d_tM^4_*}{H
\left(32 M_{\rm Pl}^2 M^4_* + 3 d_{1*} M^3_*
( 8 H M_{\rm Pl}^2+ d_{1*} M^3_*)\right)}
\\ \nn
&& - M^2_{\rm Pl} \frac{\left(32 M^4_* (6 H M_{\rm Pl}^2
+ d_{1*} M^3_*) + 6 H d_{1*} M^3_*
  (16H M_{\rm Pl}^2  +3 d_{1*} M^3_*)\right) \d_t ( d_1 M^3)_*  }
{ H d_{1*} M^3_* (4 H M_{\rm Pl}^2 + d_{1*} M^3_*)\left(32 M^2_{\rm Pl} M^4_* + 3 d_{1*} M^3_*
( 8 H M_{\rm Pl}^2+ d_{1*} M^3_*)\right)}
\end{eqnarray}
where the subscript $_*$ stands for evaluation at horizon crossing $\omega\simeq H$.

\subsubsection{3-point function in the squeezed limit}

In the de Sitter limit the computation of the 3-point function in the squeezed limit gets
largely simplified. This comes from the fact that  now (see App.~\ref{app: non linear zeta_pi})
\begin{equation}
\zeta=-H\pi+H\dot\pi \pi \, .
\end{equation}
This means that outside the horizon $\pi$ becomes constant. This simplifies the calculation in two
different ways.
On the one hand this means that there are no terms in the 3-point function of $\zeta$ which come
from the non linear relationship between $\zeta$ and $\pi$.  On the other hand, by reproducing
the argument we made in sec.~\ref{sec:explicitcalc_next to leading}, the interaction operators
with a derivative acting on each of the $\pi$'s do not contribute in the squeezed limit, and therefore
can be neglected.  Concentrating on the interaction Lagrangian with at least one $\pi$ without derivatives, we
are left with:
\begin{eqnarray}
S_3 &\supset& \int d^3x dt \left[a C_{\pi \dot{\pi} \partial^2 \pi } \pi \dot{\pi}
\partial^2 \pi + a^3 C_{\pi \dot{\pi}^2 } \pi \dot{\pi}^2\right]
\end{eqnarray}
where
\begin{eqnarray}
&&C_{\pi \dot{\pi}^2} =16 H^2 M^4_{\rm Pl}
 \frac{( 4 H M_{\rm Pl}^2 + d_1 M^3) 2 \d_tM^4
  + 2 (3 H^2 M_{\rm Pl}^2 -2  M^4 )
   \d_t(d_1 M^3)} {(4 H M_{\rm Pl}^2 + d_1 M^3)^3}\, , \\ \nn
&&C_{\pi \dot{\pi} \partial^2 \pi} =8 H^2 M^4_{\rm Pl}  \frac{ \d_t(d_1 M^3) }{(4H M_{\rm Pl}^2 +
d_1 M^3)^2}\ .
\end{eqnarray}
In the squeezed limit, we can put the first term in a form we are more familiar with by an integration
by parts:
\begin{eqnarray}
a\; C_{\pi \dot{\pi} \partial^2 \pi}(t) \pi \dot{\pi} \partial^2 \pi &=&
  -a\; C_{\pi \dot{\pi} \partial^2 \pi}(t)
  ( \partial \pi \dot{\pi} \partial \pi + \pi \partial \dot{\pi} \partial \pi) \nn\\
  &=& -a\; C_{\pi \dot{\pi} \partial^2 \pi}(t) (\pi \half \d_t (\partial \pi)^2 ) \nn\\
  &=& \half a\; \pi (\partial  \pi)^2 (H C_{\pi \dot{\pi} \partial^2 \pi}(t)
     + \dot{C}_{\pi \dot{\pi} \partial^2 \pi})\, ,
\end{eqnarray}
where in the third  passage we have neglected operators
with one derivative on each $\pi$. So in the squeezed limit we can
consider the operator $ \pi (\partial \pi)^2$ instead of $\pi
\dot{\pi} \partial^2 \pi$ upon the definition
\begin{equation}
C_{\pi (\partial\pi)^2}\equiv \frac{1}{2}(H C_{\pi \dot{\pi} \partial^2 \pi}
     + \dot{C}_{\pi \dot{\pi} \partial^2 \pi})\, .
\end{equation}
At this point it becomes immediate to obtain the result for the $\zeta$ 3-point function
using (\ref{eq:generalnongauss}) in the squeezed limit $k_1=k_L\ll k_S=k_2\simeq k_3$:
\begin{eqnarray}\label{partial 3-point2} \nn
\langle \zeta_{k_1} \zeta_{k_2} \zeta_{k_3} \rangle &=&
   -H^3 \langle \pi_{k_1} \pi_{k_2} \pi_{k_3} \rangle \\
   &\simeq& -(2\pi)^3 \delta^3( \sum_i \vec{k}_i)
  P_{k_S} P_{k_L}
  \left( \left|\pi^{cl}_{k_S}(0) \right|^2
 \frac{ C_{\pi \dot{\pi}^2}
 A_{\pi \dot{\pi}^2}
+ C_{\pi (\partial \pi)^2}
  A_{\pi (\partial \pi)^2} }{8 k_S^3 H} \right) \nonumber
\end{eqnarray}
We have already computed $A_{\pi (\partial \pi)^2}$
and $A_{\pi \dot{\pi}^2} $ in eq.(\ref{eq:2results}). Using this and
the fact that
\begin{equation}
| \pi_{k_S}(0) |^2 = \frac{1}{2 k_S^3 c_s^3 (f \tau)^2}
\end{equation}
upon substitution of eq.(\ref{eq:f and c_s}) we find a complicated expression which is
nothing but:
\begin{eqnarray}
\langle \zeta_{k_1} \zeta_{k_2} \zeta_{k_3} \rangle
&\simeq& -(2\pi)^3 \delta^3( \sum_i \vec{k}_i)
  P_{k_S} P_{k_L} n_s \ , \ \ \ k_1=k_L\ll k_S=k_2\simeq k_3 \ ,
\end{eqnarray}
with $n_s$ given by eq.(\ref{eq:tilt2}). This verifies that the consistency relation holds
also in this case.

\subsection{Verification for the Operator $\delta E^i{}_j \delta
  E^j{}_i$: the Ghost Condensate}

Now we turn to the final verification of the consistency relation, and we concentrate on the following
action:
\be
S =\int d^4x\; \sqrt{-g}\left[ \frac{M^4(t)}{2} \left(\frac{1}{N^2} - 1 \right)^2 - \frac{d_3(t)}{2} M^2(t) \delta E_{ij} \delta E^{ij}\right] \ . \label{eq:action3}
\ee
in de Sitter space.
Notice that this Lagrangian is very similar to the Lagrangian of the Ghost Condensate \cite{Arkani-Hamed:2003uy, Arkani-Hamed:2003uz}.
As explained in the former section, the $\delta E^i {}_j \delta E^j {}_i$ term introduces a spatial kinetic term for $\pi$ of the form
\begin{equation}
(\partial_i\partial_j\pi)^2 \ .
\end{equation}
Already at the level of solving for the 2-point function, this fact makes things rather more complicated,
as we will see. Furthermore, the mixing with gravity generates a term of the form
\cite{Arkani-Hamed:2003uy}
\begin{equation}
(\partial_i\pi)^2 \ ,
\end{equation}
which makes the calculation of the 2-point function even more complicated and probably undoable. However,
there is a simplifying limit we can use. This fact is carefully explained in \cite{us}, and we refer
to there for a detailed discussion. Here instead we briefly enunciate the simplifying limit. In an
inflationary background, we need to study the perturbations as they redshift
from some ultraviolet scale $\Lambda$ to
an infrared scale which is given by the Hubble constant $H$, beneath which the $\zeta$ perturbation
does not evolve anymore. Now, the terms which come from the mixing
with gravity (for example a term like $\delta N \dot\pi$, which, upon substitution of the solution
to the constraint equation $\delta N(\pi)$
becomes an operator expressible only in terms of $\pi$) are less important than the pure $\pi$ terms
at energies larger than a demixing scale $\Lambda_{\rm Mix}\simeq d_3 M^3/M_{\rm Pl}^2$ \footnote{This is a consequence of the
Goldstone equivalence
theorem applied to the field $\pi$ which non linearly realizes the time diffs.}. Now,
if $H\gg \Lambda_{\rm Mix}$, then the contribution from the mixing operators is always parametrically
suppressed. In this limit, we can therefore reintroduce the $\pi$ in the action (\ref{eq:action3})
and then put to zero all the metric fluctuations. The result we will obtain for the 3-point function
will be wrong by a small amount parameterized by $\Lambda_{\rm Mix}/H\ll 1$. Therefore, reinserting
the $\pi$ in (\ref{eq:action3}) and setting the metric fluctuations to zero,  we obtain
\be
S =\int d^4x \sqrt{-g} \left[2 M^4(t + \pi)  \dot{\pi}^2
- \frac{1}{2} d_3(t + \pi) M^2(t + \pi) \frac{1}{a^4} (\partial_i \partial_j \pi)^2\right]
\ee
Here we have neglected some higher derivative terms that do not contribute in the squeezed limit
because, as we explained in the former section, $\pi$ goes to constant outside of the horizon in the
de Sitter limit. We have also neglected a term suppressed by $H/M\ll
H/\Lambda_{\rm Mix}$. 
The action can be written as
\be \label{action3}
S =\int d^4x \sqrt{-g} \left[2 M^4 \dot{ \pi }^2 - \frac{1}{2} d_3 M^2 \frac{1}{a^4} (\partial_i \partial_j \pi )^2
- \partial_t ( d_3 M^2 )\frac{1}{2 a^4} \pi (\partial_i \partial_j \pi )^2
+ 8\dot{M} M^3 \pi \dot{\pi}^2\right] \ .
\ee

\subsubsection{2-point function and its tilt}
Solving the wave equation we find \cite{Arkani-Hamed:2003uz}:
\be
\pi_k^{cl}( \tau) = -\sqrt{\frac{\pi}{8}} \frac{H}{2 M^2} |\tau|^{3/2} H_{\frac{3}{4}}^{(1)}
\left( \frac{H k^2 \sqrt{d_3} M}{4 M^2} \tau^2 \right)\ , \label{eq:ghost wave}
\ee
where
\be
\pi_k (\tau) = \pi_k^{cl} (\tau) a_k + \pi_{-k}^{cl\; *} (\tau) a_{-k}^\dag .
\ee
The spectrum of $\zeta$ is
\be\label{eq:ghost spectrum}
P_{k} & = & \frac{\sqrt{2} \pi H^{5/2}}{\Gamma^2(1/4) k^3} \frac{1}{ M_* \left( d_{3*} M_*^2 \right)^{3/4}}\ ,
\ee
and the tilt is
\be \label{eq:ghost tilt}
n_s \simeq \frac{1}{H_*} \frac{d}{dt_*} \left(  -\log(M_*) - \frac{3}{4} \log(d_{3*} M_*^2) \right)
= -\frac{\dot{M}_*}{H_* M_*} - \frac{3 \partial_t (d_{3*} M_*^2)}{4 H_* d_{3*} M_*^2}\ .
\ee

\subsubsection{3-point function in the squeezed limit}

From eq.(\ref{action3}), we see that we have to compute the contribution of two operators in
the squeezed limit ($k_1=k_L\rightarrow 0$). Let us start with  $\pi (\partial_i \partial_j \pi )^2$:
\be
\langle \pi_{k_1}\pi_{k_2}\pi_{k_3} \rangle_{\pi (\partial_i \partial_j \pi )^2} & = &- i \partial_t ( d_3 M^2 )
(2 \pi)^3 \delta^3 \big( \sum_i k_i \big) \pi^{cl}_{k_1}(0) \pi^{cl}_{k_2}(0) \pi^{cl}_{k_3}(0)  \\
& & \times \int d \tau \ \! \pi^{cl*}_{k_1}(\tau) \pi^{cl*}_{k_2}(\tau) \pi^{cl*}_{k_3}(\tau) (k_2 \cdot k_3)^2  + c.c. \nonumber \\
& = & -i \partial_t ( d_3 M^2 ) (2 \pi)^3 \delta^3 \big( \sum_i k_i \big)
\frac{-i H^{3/4} \Gamma^3(3/4) }{(d_3 M^2)^{9/8} (\sqrt{2}\pi M)^{3/2} (k_1 k_2 k_3)^{3/2}}\nonumber \\
& & \times \int d \tau \ \! \pi^{cl*}_{k_1}(\tau) \pi^{cl*}_{k_2}(\tau) \pi^{cl*}_{k_3}(\tau) (k_2 \cdot k_3)^2 + c.c. \nonumber \\
& \simeq & -i \partial_t ( d_3 M^2 ) (2 \pi)^3 \delta^3 \big( \sum_i k_i \big)
\frac{-i H^{3/4} \Gamma^3(3/4) k_S}{(d_3 M^2)^{9/8} (\sqrt{2}\pi M)^{3/2} k_L^{3/2}} \nonumber \\
& & \times \int d \tau \left( \pi^{cl*}_{k_S}(\tau) \right)^2
\left( -i  \frac{1}{2M^2} \left( \frac{8 H M^3}{d_3^{3/2} } \right)^{1/4} \frac{\Gamma(3/4) }{k_L^{3/2} \sqrt{\pi}}  \right) + c.c. \nonumber \\
& = & i \partial_t ( d_3 M^2 ) (2 \pi)^3 \delta^3 \big( \sum_i k_i \big)
\frac{H \Gamma^4(3/4) k_S}{(d_3 M^2)^{3/2} 2M^2 \pi^2 k_L^3}
\int d \tau \left( \pi^{cl*}_{k_S}(\tau) \right)^2 + c.c.\ . \nn
\ee
where in the third passage we have used the small argument expansion of the wavefunction
(\ref{eq:ghost wave}). This
approximation is justified for the same reason as we argued in sec.
\ref{sec:explicitcalc_next to leading}.
Now the integral is
\be
I_{\pi (\partial_i \partial_j \pi )^2} & = & \int d \tau \left( \pi^{cl*}_{k_S}(\tau) \right)^2  \nonumber \\
& = & \frac{\pi}{32} \frac{H^2}{M^4} \left[\int d \tau \ \! \tau^3
\left( H_{3/4}^{(1)} \left( \frac{H k^2_S \sqrt{d_3} M}{4 M^2} \tau^2 \right) \right)^2\right]^* \nonumber \\
& = & \frac{\pi}{32} \frac{H^2}{M^4} \frac{16 M^2}{d_3 H^2 k_S^4} \frac{1}{2} \left[\int d x \ \! x
\left( H_{3/4}^{(1)} (x) \right)^2\right]^* \nonumber \\
& = & \frac{\pi}{4} \frac{1}{d_3 M^2 k_S^4} \left( -\frac{3}{2 \pi} (1 - i) \right) \nonumber \\
& = & -\frac{3}{8} \frac{1}{d_3 M^2 k_S^4} (1 - i)
\ee
via the substitution $x = \frac{H k_S^2 \sqrt{d_3} }{4 M} \tau^2$. We have also used that
\be
\Gamma(3/4) = \frac{\sqrt{2} \pi }{\Gamma(1/4)} \ .
\ee
So, the contribution from the operator $\pi (\partial_i \partial_j \pi
)^2$ for $ k_1=k_L\ll k_S= k_2\simeq k_3$ is
\be
\langle \pi_{k_1}\pi_{k_2}\pi_{k_3} \rangle_{\pi (\partial_i \partial_j \pi )^2} =- 6 \pi \delta^3 \big( \sum_i k_i \big)
\partial_t ( d_3 M^2 ) \frac{H \Gamma^4(3/4)}{2M^2 (d_3 M^2)^{5/2}
  k_S^3 k_L^3}\ , 
\ee
Next let us compute the contribution from the operator $\pi
\dot{\pi}^2$ in the same limit.  This is very similar,
we just have a slightly different integral
\be
\langle \pi_{k_1}\pi_{k_2}\pi_{k_3} \rangle_{\pi\dot{\pi}^2} & = & i 16 \dot{M} M^3
(2 \pi)^3 \delta^3 \big( \sum_i k_i \big) \pi^{cl}_{k_1}(0)
\pi^{cl}_{k_2}(0) \pi^{cl}_{k_3}(0) \nonumber  \\
& & \times \int \frac{d \tau}{(H \tau)^2}  \pi^{cl*}_{k_1}(\tau) \partial_\tau \pi^{cl*}_{k_2}(\tau) \partial_\tau \pi^{cl*}_{k_3}(\tau)  + c.c. \nonumber \\
& \simeq & - i 16 \dot{M} M^3 (2 \pi)^3 \delta^3 \left( \sum_i k_i
\right) \frac{\Gamma^4(3/4) }{2 H M^2 (d_3 M^2)^{3/2} \pi^2 k_L^3 k_S^3}
\int d \tau \tau^{-2} \left( \partial_\tau \pi^{cl*}_{k_S}(\tau)
\right)^2 \ , \nonumber \\
&& 
\ee
where again we have used the long wavelength expansion for $\pi^{cl}_{k_1}$ in (\ref{eq:ghost wave}).
The remaining integral gives:
\be
I_{\pi\dot{\pi}^2} & = & \int d \tau\; \tau^{-2} \left( \partial_\tau \pi^{cl*}_{k_S}(\tau) \right)^2  \nonumber \\
& = & \frac{\pi H^2}{16 M^4} \left[\int dx\; x \left( H_{-1/4}(x) \right)^2 \right]^*\nonumber \\
& = & \frac{H^2}{32 M^4} (1 - i)\ ,
\ee
so that the result for this term after adding the complex conjugate is
\be
\langle \pi_{k_1}\pi_{k_2}\pi_{k_3} \rangle_{\pi \dot\pi^2}
=-4\pi\delta^3 \big( \sum_i k_i \big) \frac{ \dot{M} H
  \Gamma^4(3/4)}{M^3 (d_3 M^2)^{3/2} k_L^3 k_S^3}\ .
\ee

Thus the full result in the squeezed limit is ($ k_1=k_L\ll k_S= k_2\simeq k_3$)
\be
\langle \pi_{k_1} \pi_{k_2} \pi_{k_3} \rangle
=  -2 \pi \delta^3 \big( \sum_i k_i \big) \left(\frac{ 3 \partial_t ( d_3 M^2 ) }{d_3 M^2} + 4 \frac{\dot{M}}{M} \right)
\frac{H \Gamma^4(3/4)}{2 M^2 (d_3 M^2)^{3/2} k_1^3 k_2^3} ,
\ee
converting this using $\zeta_k = -H \pi_k$ gives
\be
\langle \zeta_{k_1} \zeta_{k_2} \zeta_{k_3} \rangle
= 2 \pi \delta^3 \big( \sum_i k_i \big) \left( 4 \frac{\dot{M}}{M} + 3 \frac{ \partial_t ( d_3 M^2 ) }{d_3 M^2} \right)
\frac{H^4 \Gamma^4 (3/4)}{2M^2 (d_3 M^2)^{3/2} k_L^3 k_S^3}  ,
\ee
where $k_L=k_1$ and $k_S= k_2\simeq k_3$.
The consistency relation is
\be \label{eq:help}
\lim_{k_1 \to 0} \langle \zeta_{k_1} \zeta_{k_2} \zeta_{k_3} \rangle
= -(2 \pi)^3 \delta^3 \big( \sum_i k_i \big) n_s P_{k_L} P_{k_S} ;
\ee
so, in order to verify it,  we need to compute the right hand side of this equation
using eq.s (\ref{eq:ghost spectrum}) and (\ref{eq:ghost tilt}):
\be
- (2 \pi)^3 \delta^3 \big( \sum_i k_i \big)  n_s P_{k_L} P_{k_S} & = &
(2 \pi)^3 \delta^3 \big( \sum_i k_i \big) \left( \frac{\dot{M}}{H M} +\frac{3 \partial_t (d_3 M^2)}{4 H d_3 M^2} \right)
\frac{(2 \pi)^2 H^5}{2M^2 (d_3 M^2)^{3/2} \Gamma^4(1/4) k_S^3 k_L^3} \nonumber \\
& = &  2 \pi \delta^3 \big( \sum_i k_i \big)  \left( 4 \frac{\dot{M}}{M} + 3 \frac{\partial_t (d_3 M^2)}{d_3 M^2} \right)
\frac{H^4 \Gamma^4(3/4)}{2M^2 (d_3 M^2)^{3/2} k_S^3 k_L^3} .
\ee
We see that the consistency relation is again satisfied.

\section{Conclusions\label{sec:conclusions}}
Observation of the non-Gaussian component of the CMB is going to improve rapidly in the next
few years with the launch of the Planck satellite. This will reduce the current limit from WMAP 3yr data
\cite{Creminelli:2006rz} by a factor of around 6 \cite{Babich:2004yc}. While standard slow
roll inflation predicts a level on non-Gaussianity far beyond current sensitivity, there are many models
of inflation which predict a larger level on non-Gaussianity and that are already beginning to be
constrained by current observations.

For this reason, we consider it very important to understand the properties of the non-Gaussian signal
coming from inflation. Along this line of reasoning,
in \cite{Maldacena:2002vr, Creminelli:2004yq} it was pointed out
that in all models of inflation with only one relevant degree of freedom, which acts as the clock of the
system, the 3-point function
in a particular geometrical limit, the squeezed triangle limit, should
follow a consistency relation:
\begin{equation}\label{eq:cc2}
\lim_{k_1 \to 0} \langle\zeta_{\vec k_1} \zeta_{\vec k_2} \zeta_{\vec
  k_3}\rangle = - (2 \pi)^3 \delta^3(\sum_i \vec k_i) P_{k_1} P_{k_3}
\frac{d \log k_3^3 P_{k_3}}{d \log k_3} \;,
\end{equation}
This consistency relation involves a level of non-Gaussianity too small to be detectable by foreseeable
experiments. However, it is still very important for the following reason: if we detect
some signal in the squeezed triangle configuration, it will mean that all the single clock inflationary
model will be ruled out.

Since this is a very powerful statement, we think it is very important to be sure that the consistency
relation (\ref{eq:cc2}) is true. The purpose of the present paper has been to further settle this issue.
In the first part of the paper, we have made more explicit and rigorous
the proof that was already present in
\cite{Maldacena:2002vr, Creminelli:2004yq}. Still this proof turns out to be rather implicit, even though
physically clear. For this reason, we found it useful to prove the consistency relation in a completely
orthogonal way, i.e. through a direct check of all possible single field models. Clearly, this task
seems at first rather difficult and ill defined. However, we have been able to do
this because we could exploit a
recently developed effective field theory for inflation \cite{Creminelli:2006xe, us} which
completely describes the fluctuations around an inflationary background under the very general
hypothesis that there is only one dynamical system and that
this spontaneously breaks time diffeomorphisms.

The calculations are a bit convoluted, even though we have been able to use some simplifying tricks, and
for some complicated models we have performed the study only in some simple limit. However,
the result is very immediate: the consistency relation of the 3-point function of all single clock models {\it does} hold, and hopefully this will help us learn something new from the upcoming
cosmological experiments.

\section*{Acknowledgments}
It is a pleasure to thank Paolo Creminelli for initial collaboration in this project, in particular
in obtaining the results of sec.~\ref{sec: formal proof}. We would like to thank also Nima Arkani-Hamed
and Matias Zaldarriaga for continuous encouragement.  A. Liam Fitzpatrick is supported by an NSF fellowship. Jared Kaplan is supported by a Hertz fellowship and an NSF fellowship.

\appendix

\section*{Appendix}

\section{Matching our Unitary Gauge Lagrangian to a general
$P(X,\phi)$ Lagrangian \label{app: matching theories}}


As one should have expected, our effective Lagrangian
reproduces the result of the widely used k-inflation Lagrangian
\cite{Armendariz-Picon:1999rj}, as we are now going to verify.

The Lagrangian of k-inflation can be written in the form: \be
S_{\rm k-inf} = \int d^4 x \sqrt{-g} P(X,\phi)\ ,
\label{eq:kinflag}   \ee where
$X=-g^{\mu\nu}\partial_\mu\phi\partial_\nu\phi$. To match our
effective Lagrangian to the k-inflation Lagrangian, we just
write the k-inflation Lagrangian in unitary gauge
$\phi(\vec{x},t)=\phi_0(t)$. The $X$ variable becomes:
\begin{eqnarray}
X  = \frac{\dot{\phi}_0(t)^2}{N^2}
\end{eqnarray}
Obviously, the factor $1/N^2$ can only enter the unitary gauge
Lagrangian in terms with $X_0(t)=\dot\phi_0^2$.  For this reason, an expansion
of $S_{\rm k-inf}$ in powers of $1/N^2 - 1$ is the same as an
expansion in powers of $X-\dot\phi_0^2$. We obtain:
 \be
c_n(t)\; M^4(t) &=& \dot{\phi}_0(t)^{2n} \frac{\d^n}{\d X^n}
 P(X,\phi)\big\vert_{0} \  .
 \ee
where in our Lagrangian $c_2(t)\equiv 1$, and $c_1(t)=-M^2_{\rm Pl}\dot H$
and $c_0(t)=-M^2_{\rm Pl}(3H^2+2\dot H)$.

 In particular, we find that the speed of sound is
\be c_s^{-2} &=& 1-\frac{2 M(t)^4}{\dot H M_{\rm Pl}^2} \\ &=&
 1+2 \dot{\phi}_0^2 \frac{\d^2 P}{\d X^2}\Big\vert_{0} \left( \frac{\d P}{\d X}\Big\vert_0\right)^{-1}
\label{eq:cs}\ee

In general the use of a polynomial as $P(X,\phi)$ is not
consistent in a regime of effective field theory, because the
function $P$ involves many non-renormalizable terms. However, it
is possible that some UV complete theory can induce a low energy
effective theory with the structure of k-inflation, where a
precise relationship between all the non-renormalizable operators
is understood. This is the case for DBI inflation, a scenario in
which the inflaton represents the coordinate of a brane traversing
 an extra dimensional warped geometry. Due to the causal
speed limit in the extra dimension, the inflaton cannot roll
arbitrarily fast, and in fact experiences a relativistic drag that
admits slow roll behavior despite a steep potential
\cite{Alishahiha:2004eh}. This effect can also be understood in
terms of the purely four dimensional CFT dual because as the
inflaton rolls, new light scalar modes become accessible.
Integrating out these modes induces a frictional force
encapsulated by the low energy Lagrangian \cite{Alishahiha:2004eh}
\be S_{\rm DBI} &=& \int d^4 x \sqrt{-g}
\left[-f^{-1}(\phi)\sqrt{1-f(\phi)X} - V(\phi) \right]. \ee Notice
the resemblance of this Lagrangian to that of a relativistic point
particle in an external potential. Indeed, following this analogy
we see that without the potential the DBI Lagrangian possesses a
residual 5D spacetime symmetry corresponding to arbitrary
``boosts'' in the $\phi$ direction. Such a transformation rotates
the $d\phi$ direction into the $dt$ time direction, and so it acts
nonlinearly on derivatives of the inflaton \be
\dot{\phi} &\rightarrow& \dot{\phi} + \eta(f^{-1/2}- f^{1/2}\dot{\phi}^2), \\
dt &\rightarrow& dt(1+ \eta f^{1/2}\dot{\phi}), \ee where $\eta$
is an infinitesimal boost parameter.  In the UV, this symmetry
comes from the isometries of ${\rm AdS}_4$, or, dually, from
superconformal invariance of $\mathcal{N}=4$ SYM
\cite{Maldacena:1997re}.

From the the nonlinear form of this boost, it is clear that
coefficients of higher order terms in $\dot{\phi}$ are uniquely
set by those of lower order terms. Consequently, the square root
form of the DBI action is uniquely set by this symmetry, and
moreover we find that
\begin{equation}
c_n(t)M^4(t) \sim \frac{\dot{\phi}_c^{2n} f(\phi_c)^{n-1}}{(1- f
\dot{\phi}_c^2)^{n-1/2}} \, , \ \ \ n\geq 2 \ , \label{dbimatch1}
\end{equation}
where we have neglected some numerical coefficients. Since the FRW
equations set the value of $c(t)$ and $\Lambda(t)$, and the speed
of sound sets the ratio of $M(t)^4$, as usual we can rewrite
$\dot{\phi}_0^2$ and $f(\phi_0)$ in terms of $H$, $c_s$, and slow
roll parameters, giving \be \dot{\phi}_0^2 &=& 2 \epsilon H^2 M^2_{\rm Pl} c_s , \\
f(\phi_0) &=& \frac{c_s}{2 \epsilon H^2 M^2_{\rm Pl}}
\left(\frac{1}{c_s^2}-1\right) . \ee  where $\epsilon=-\dot H/H$. Plugging these expressions
back into (\ref{dbimatch1}), we obtain
\begin{equation}
c_n(t) M(t)^4 \sim \epsilon H^2 M^2_{\rm Pl}
\left(\frac{1}{c_s^2}-1\right)^{n-1}\ , \ \ \ n\geq 2\ ,
\end{equation}
so we see that the symmetries of the DBI action demand a very
particular form for the the coefficients of our effective
Lagrangian. In particular, we can also see that, in the limit of
$c_s\ll 1$, the symmetries impose an hierarchy between the
$c_n$'s.

\section{Non Linear Relation between $\pi$ and $\zeta$\label{app: non linear zeta_pi}}

We will work out the relationship between the $\pi$ gauge
(\ref{eq: pi gauge}) and the $\zeta$ gauge (\ref{zetagauge})
using the results of \cite{Maldacena:2002vr}.  Since $\zeta$ is
the relevant variable which is constant outside the horizon, we
need to know $\zeta$ in terms of $\pi$ in order to determine the
observational consequences of our effective Lagrangian.  Let us remind
ourselves which gauge properly defines the $\zeta$ variable. This is not the spatially
flat slicing we used in deriving the next to leading order Lagrangian in $\pi$ eq. (\ref{eq: pi gauge}), but it
is rather a gauge fixed version of the unitary gauge we used to build our effective Lagrangian.
In $\zeta$ gauge time diffs are fixed by imposing precisely that $\pi=0$, while spatial diffs
are fixed by requiring the spatial metric to be isotropic. In other words, the $\zeta$ gauge is defined
by the condition:
\begin{equation}\label{eq:zetagauge}
\pi=0\; , \ \ \ \ \ \hat{g}_{ij}=a^2e^{2\zeta}\delta_{ij} \ .
\end{equation}
If we denote by $\tilde t$ the time coordinate in the $\pi$ gauge, and by $t$ the time in
the $\zeta$ gauge, we have to perform a time reparametrization of the form $\tilde t= t+T(x)$ to
go from $\pi$ gauge to the $\zeta$ gauge. We have
then to solve the following equation:
\begin{equation}
\pi(\tilde x)\rightarrow \pi_\zeta(x(\tilde x))=\pi(\tilde x)+T(x(\tilde x))=0
\end{equation}
where in the second passage we have used the fact that the $\pi_\zeta$, {\it i.e.} the $\pi$ in $\zeta$ gauge,
is zero.
This implies that \be T(x) = - \pi(x) + \dot{\pi}(x) \pi(x) \ee to
second order. Maldacena \cite{Maldacena:2002vr} worked out $\zeta$
in terms of $T(x)$; to quadratic order he found \be
\label{ZetaPiRelation}
\zeta & = & H T + \frac{1}{2} \dot{H} T^2 + \alpha(T) \\
& = & -H \pi + H \dot{\pi} \pi + \frac{1}{2} \dot{H} \pi^2 +
\alpha(T(\pi)) \ee where $\alpha$ is determined by solving for the
additional spatial diffeomorphism needed to maintain the $\zeta$
gauge condition. Although the $\alpha$ term is of second order in $T$, it
only contains higher derivative terms that vanish outside the
horizon, so it is irrelevant for our calculation.

\section{Wave Equation at First Order in Slow Roll for the case $\omega^2=c_s^2k^2$
\label{app: waveeq}}

Using the conformal time $d\tau = dt/a$, the quadratic action
$S_2[\pi]$ is \be S_2 &=& \frac{1}{2} \int d^3 x d\tau (a^2 f^2)
 \left[ \frac{\pi'^2}{a^2} - \frac{c_s^2}{a^2}(\partial \pi)^2 -m^2
\pi^2 \right]+\ldots, \\ f^2 &=& -\frac{2 a^2 \dot{H}M^2_{\rm Pl}}{c_s^2},\\
m^2 &=& 3 \dot{H}, \ee where dots and primes denote derivatives
with respect to $t$ and $\tau$, respectively, and $\tau$ runs from
$-\infty$ to $0$. Applying the field redefinition \be \pi &=&
f^{-1}\sigma, \label{eq:sigmadef} \ee and going to Fourier space,
we obtain a wave equation for $\sigma_k(\tau)$, \be \sigma_k'' +
c_s^2
k^2 \sigma_k &=& \left(-a^2 m^2 + \frac{f''}{f}\right)\sigma_k \\
&=& 2a^2 H^2 \left(1-\frac{\epsilon}{2}+\frac{3 \eta}{4} -
\frac{3s}{2}\right)\sigma_k. \ee Solving the wave equation, and imposing to be in the Minkowski
vacuum at early times, we
find that classically \be \pi_k^{\rm cl}(\tau) &\sim&
(-\tau)^{3/2} H_\nu^{(1)}(-c_s k \tau (1+s)),\\
\nu &=& \frac{3}{2} + \epsilon + \frac{\eta}{2} + \frac{s}{2}. \ee
We determine the overall normalization of this function when we
quantize $\pi$.

We promote $\pi_k$ to a field operator \be \pi_k(t) &=& \pi_k^{
cl}(t) a_k + \pi_{-k}^{{cl}*}(t) a_{-k}^\dagger, \ee where
$\pi_k^{cl}$ satisfies the wave equation and has normalization
set by the canonical commutation relation \be [a_k,a_{k'}^\dagger]
= (2 \pi)^3\delta^{(3)}(k-k'), \ee yielding
\begin{eqnarray}
\pi^{cl}_k(\tau) &=& -\sqrt{\frac{\pi}{8\epsilon}} \frac{c_s}{M_{\rm
Pl}} (-\tau)^{3/2}(1-\epsilon_* + s_*/2)
e^{i\pi(\epsilon_*/2 + \eta_*/4)} H_\nu^{(1)}(-c_s k \tau (1+s)),\\
\nu &=& \frac{3}{2} + \epsilon + \frac{\eta}{2} + \frac{s}{2}.
\label{piwavefinc}
\end{eqnarray}
We follow \cite{Chen:2006nt}, and define $\epsilon_*$, $\eta_*$
and $s_*$ as the slow roll parameters evaluated at the time at
which $K_1 = k_1 + k_2 + k_3$ exits the horizon, so that $c_s K_1 =
a H$.
 One can explicitly calculate the tilt
of the power spectrum for $\zeta$ by expanding $|\zeta_k|^2=|H
\pi_k(\tau)|^2$ in the late time limit, $k\rightarrow 0$.  In this
regime, the Hankel function behaves as $H_\nu(k) \propto
k^{-\nu}$, so the two-point correlator behaves as $k^{-3 - 2
\epsilon - \eta - s} \equiv k^{-3 + n_s}$.  This shows very
explicitly that
\begin{equation}
n_s = -2 \epsilon - \eta - s \ .
\end{equation}

Setting to zero the slow roll parameters, we recover the de Sitter
mode \be \pi_k^{cl}(\tau) &=& \frac{i}{2M_{\rm Pl}k^{3/2}
\sqrt{c_s \epsilon}}(1+i c_s k \tau) e^{-i c_s k \tau}. \ee

\subsection{Explicit Long Wavelength Behavior of $\dot{\zeta}$ \label{sec:long wave limit}}

One of the major advantages of the $\zeta$ variable is
that it goes exponentially fast to a constant when it is outside of the
horizon. In conformal time, this exponential speed turns into a
power law behavior. This is clearly manifest in the classical
$\zeta$ modes at leading order in the slow roll parameters, where
$\zeta = -i \frac{H}{2 M_{\rm Pl} \sqrt{\epsilon c_s k^3}} (1+ i
c_s k \tau)e^{-i c_s k \tau}$, but not at higher
orders. However, at next to leading order, in the long wavelength
limit, the classical modes can be written in closed form.

To begin, we expand $H_\nu^{(1)}(x) = J_{\nu}(x)+ i Y_\nu(x)$
around $\nu = 3/2$ and $x=0$:
\begin{equation}
H_\nu^{(1)}(x) =
   - \sqrt{\frac{2}{\pi}} i \frac{(1-ix)e^{ix}}{x^{3/2}}
  \left( 1 - (\nu - \frac{3}{2})
  \left[ \gamma - 2 + \log 2 + \log x \right] \right)
\end{equation}
The constant $\gamma = 0.577\dots$ is the Euler-Mascheroni
constant. Inserting this in our expression for the classical $\pi$
modes in equation (\ref{piwavefinc}), we obtain, dropping the irrelevant constant phase:
\begin{equation}
\pi^{cl}_k(\tau) =
  \frac{i}{2} \frac{ \epsilon^{-1/2} c_s^{-1/2} }{M_{\rm Pl} k^{3/2}}
  ( 1 + \frac{1}{2}(c_s k \tau (1+s))^2) (1 - \epsilon_* +\frac{s_*}{2}-\frac{3s}{2})
 ( 1 - ( \epsilon + \frac{\eta}{2} + \frac{s}{2} )
( \gamma -2 + \log 2 + \log (-c_s k \tau (1+s)) ))
\end{equation}
The additional time-dependence that appears at next-to-leading
order in slow-roll is the $\log \tau$ component as well as the
time-dependence of the parameters $c_s,\epsilon, \eta, s$, themselves. When
we differentiate with respect to time, the derivatives of
slow-roll parameters will be present at next to leading order only
for parameters that were already present in the leading order
modes.  With this in mind,
 it is straightforward to calculate that
\begin{equation}
  \dot{\zeta}_k =
   i \frac{H^2 \tau^2}{2 M_{\rm Pl}} \sqrt{ \frac{c_s^3 k}{\epsilon} }
   \left( 1 + (2\epsilon-\epsilon_*+ \eta + \frac{s}{2}+\frac{s_*}{2}) + (\epsilon +\frac{\eta}{2} + \frac{s}{2} )
   (-\gamma - \log (-2 k \tau c_s) ) \right)
\end{equation}
Individual terms in $H \dot{\pi}$ or $\dot{H} \pi$ proportional to
negative powers of $k$ have cancelled out.

\section{Proof that $\zeta$ is Constant Outside the Horizon \label{app: zeta cosntant}}

In this Appendix we generalize Maldacena's proof
\cite{Maldacena:2002vr} (see also \cite{Salopek:1990jq}) that $\zeta$ is constant outside the
horizon to a completely general model of inflation with one degree
of freedom.  The idea of the proof is simple -- we will expand the
action to first order in \emph{derivatives} of the dynamical field
$\zeta$, but to all orders in $\zeta$ itself, without any other
approximations.  We will show that to this order, the action is a
total derivative, so the $\zeta$ action must begin at second order
in derivatives.  Thus $\dot{\zeta} = 0$ is always a solution of
the equations of motion when we neglect spatial derivatives. Moreover,
we shall briefly show that the constant solution is an attractor, in the sense
that solutions in its neighborhood approach it exponentially fast in time.
As we show in the main text, the solution to the linearized $\zeta$ equation which
deep inside the horizon is in the Minkowski vacuum, at late times
asymptotes exponentially to the $\zeta=\;$const solution \footnote{The fact that one
of the two solutions to the linearized equation is exponentially damped outside of the horizon
is very general: in the $\pi$ equation, it comes
from the friction term $3H\dot\pi$ where $\pi=-\zeta/H$ at linear level.}. Since we are always
in the perturbative regime this tells us that we are in the basin of attraction of the
constant solution, and that therefore $\zeta$ is constant outside of the horizon at
non linear level.

The first step of the proof is to show self-consistently that
$\delta N, \hat\nabla_i N_j = \mathcal{O}(\partial_\mu \zeta)$ \footnote{At linear order
the solutions for $N$ and $N^i$ in the gauge of $\zeta$ are given by:  
\begin{eqnarray}\label{constr_zeta}
&&N-1= \frac{A_1}{C}\dot\zeta- \frac{B_1}{C} (\partial^2\zeta/a^2)\ \
, \ \ \partial_i N^i= \left(\frac{A_2}{C}\dot\zeta+\frac{B_2}{C}(\partial^2\zeta/a^2)\right)\ ,
\end{eqnarray}
where, up to small terms suppressed by $M/M_{\rm Pl}\ll 1$ or $H/M\ll 1$, the 
coefficients are:
\begin{eqnarray}
&&A_1= 4 M_{\rm Pl}^2\left(d_1 M^3-4 M_{\rm Pl}^2 H\right) \ \ , \ \ B_1= 8 M_{\rm Pl}^2 \left(d_2+d_3\right)M^2\ , \\
&&A_2= 16 M_{\rm Pl}^2 \left(-2 M^4+M^2_{\rm Pl}\dot H\right) \ \ , \ \ B_2=4 M_{\rm Pl}^2\left(d_1 M^3+4 M^2_{\rm Pl} H\right)\ , \\
&&C=\left(d_1^2 -16 \left(d_2+d_3\right)\right)M^6-16 M_{\rm Pl}^4H^2 \ .
\end{eqnarray}
The linear solution for $\zeta$ vanishes outside of the horizon quickly enough so that $\delta N$ and $N^i$ go to zero in the same limit.
 }.
This makes sense intuitively --
since $N$ and $\hat \nabla_j N_i$ are constrained variables, while $\zeta$ is the only dynamical
degree of freedom, if derivatives of $\zeta$ were to vanish,
then we would simply have a pure FRW cosmology up to a trivial rescaling of
the coordinates, and $\delta N$
and $\hat\nabla_i N_j$ would also vanish.  Thus they must be
proportional to derivatives of $\zeta$.
If we assume that this is the case,
then we only need the action to quadratic order
in $\delta N$ and $\hat\nabla_i N_j$.  Thus for
our purposes, the relevant piece of the action is
\be
S = \int d^3 x dt \sqrt{\hat{g}} \left[N \frac{M_{\rm Pl}^2}{2} R^{(3)}
+ \frac{1}{N} \frac{M_{\rm Pl}^2}{2} \left( E_{ij} E^{ij} - {E^i_i}^2 \right) - \frac{1}{N} M_{\rm Pl}^2 \dot{H}
- N M_{\rm Pl}^2 (3 H^2 + \dot{H}) \right. \nonumber \\
+ \left. \frac{1}{2} N M(t)^4 \left( \frac{1}{N^2} -1 \right)^2 +
\frac{d_1}{2} M(t)^3 N \delta N \delta E_i^i-\frac{d_2(t)}{2}\;
M(t)^2\delta E^i {}_i {}^2 -\frac{d_3(t)}{2}\; M(t)^2 \delta E^i
{}_j \delta E^j {}_i \right] . \ee Now since $N$ is simply a
Lagrange multiplier, we can vary the action with respect to it to
obtain its (algebraic) equation of motion \be 2 \delta N \left( 3
M_{\rm Pl}^2 H^2 + M_{\rm Pl}^2 \dot{H} - 2M^4 \right) =
\dot{\zeta} \left( 6 M_{\rm Pl}^2 H + \frac{3}{2} d_1 M^3 \right)
+ \hat\nabla_i N^i \left( -2 M_{\rm Pl}^2 H - \frac{d_1}{2} M^3
\right) \ee where we have used the fact that \be E_{ij} E^{ij} -
{E^i_i}^2 = -6 \left( H + \dot{\zeta} \right)^2 + 4 \left( H +
\dot{\zeta} \right) \hat\nabla_i N^i+{\cal{O}}\left((\hat\nabla_i
N^i)^2\right) . \ee Analogously for $\hat\nabla_jN_i$ we obtain:
\begin{eqnarray}
&&\frac{1}{2} M_{\rm Pl}^2 \left[ 4 \hat\nabla_i \left( \dot{\zeta} - H \delta
N  \right)  +
\hat\nabla_j \hat\nabla^j N_i +
\hat\nabla_j  \hat\nabla_i N^j - 2 \hat\nabla_i \hat\nabla_j N^j \right] -
\frac{d_1}{2} M(t)^3 \hat\nabla_i \delta N  + \\ \nn &&
d_2 M(t)^2 \left( 3 \hat\nabla_i  \dot\zeta -
\hat\nabla_i \hat\nabla_j N^j \right) +
 \frac{d_3}{2} M(t)^2 \left( 2 \hat\nabla_i\dot{\zeta} -
\hat\nabla_j  \hat\nabla^j N_i  -
\hat\nabla_j  \hat\nabla_i N^j \right)
= 0
\end{eqnarray}
We see that it is self-consistent to assume that
$\delta N, \hat\nabla_i N_j = \mathcal{O}(\partial_\mu \zeta)$, which
justifies our neglect of higher powers of these parameters.

Now let us expand the action to linear order in $\partial_\mu \zeta$.
For this purpose, note that $R^{(3)}$ is of quadratic order,
so we can neglect it.  We find
\be
S  =  \int d^3x dt\; a(t)^3 e^{3 \zeta} M_{\rm Pl}^2 \left[
\frac{1}{2} (1 - \delta N) \left(-6 \left( H^2 + 2 H \dot{\zeta} \right)
+ 4 H \hat\nabla_i N^i \right) - (1 - \delta N) \dot{H} \right. \nonumber \\
\left. - (1 + \delta N) \left( 3 H^2 + \dot{H} \right) \right] ,
\ee
if we simplify and integrate the $\hat\nabla_i N^i$ term by parts, we find
\be
S & = & \int d^3x dt\; a(t)^3 e^{3 \zeta} M_{\rm Pl}^2 \left[
- 6 H^2 - 2 \dot{H} - 6 H \dot{\zeta} \right] \nonumber \\
& = & \int d^3x dt \frac{d}{dt} \left[- a(t)^3 e^{3 \zeta} 2 H M^2_{\rm Pl}\right] .
\ee
Note that the term linear in $\delta N$ dropped out of the action.
This is not surprising -- it is due to the fact that
$\frac{\partial L}{\partial (\delta N)} = 0$ is satisfied to
zeroth order in derivatives, simply because the metric satisfies
Einstein's equations, and we have neglected terms at
second order in derivatives.

As claimed, the action for $\zeta$ begins at second order in
derivatives for all models of inflation based on a single degree
of freedom. This means that $\zeta=\;$const is always a solution
of the equations of motion when we neglect gradients. However, the
solution we are interested in, which asymptotes to the Minkowski
vacuum deep inside the horizon, is not exactly constant. At linear
level, the time dependent component goes to zero exponentially
fast in time. We want to verify this at non linear level, by
showing that in fact the $\zeta=\;$const solution is still an
attractor.

This can be done rather quickly by the following reasoning. Let us
consider a small perturbation around a solution
$\zeta=\zeta_0=\;$const:
\begin{equation}
\zeta=\zeta_0+\psi\ ,
\end{equation}
and let us ask what is the linearized equation of motion for
$\psi$. By the same definition of $\zeta$
(eq.(\ref{eq:zetagauge})), a constant component $\zeta_0$ can be
reabsorbed in the definition of $a(t)$: $\tilde
a(t)=a(t)\;e^{\zeta_0}$. This implies that the linearized equation
of motion of $\psi$ in a $\zeta_0$ background has to be the same
as the one of $\zeta$, with $a(t)$ replaced with $\tilde a(t)$.
Therefore the solution for $\psi$ is the same as for the
linearized $\zeta$: out of the horizon it either decays
exponentially fast in cosmic time, or it goes to a constant (and
in this case it amounts at just a redefinition of $\zeta_0$).

We therefore conclude that outside of the horizon the solution for
$\zeta$ quickly converges to a constant and $N$ and $N^i$ tend to their unperturbed values. At this point, via the
argument of Sec. \ref{sec: formal proof}, we deduce that the
consistency relation between the three point function of $\zeta$
and the tilt of the spectrum must hold for the models with only
one relevant degree of freedom.




\section{Reintroducing the $\pi$ in $\delta E^i {}_i$ \label{App: deltaNdeltaE}}

Since the reintroduction of the $\pi$ field in $\delta E^i {}_i$ is not entirely trivial,
 in this appendix we perform an explicit calculation. In the spatially flat gauge
(\ref{eq: pi gauge}) in which we are working, the trace of $\delta E_{ij}$ can be written as:
\begin{equation}
\delta E_i^i = \half \hat g^{ij} \partial_t\hat g_{ij} - \partial_i N^i- 3 H
\end{equation}
The reintroduction of the $\pi$ field follows the same steps we illustrate of
\cite{us}. What makes this case slightly more complicated than the case
explicitly illustrated in \cite{us} is the fact that the transformation of
$\hat g_{ij}$ and $N_i$ under time diffeomorphisms are more complicated than the one of $N$.
The following equations
are true:
\be
 g^{0i} &=& \frac{N^i}{N^2}  \nn\\
-  g^{00} &=& \frac{1}{N^2}  \, .
\en
where $g_{\mu\nu}$ is the full 4d metric.
Taking variations of $g^{0i}$, we find
\be
 \delta g^{0i} &=& \frac{1}{N^2} \delta N^i + N^i \delta(\frac{1}{N^2})
  + \delta N^i \delta (\frac{1}{N^2})\, .
\en
Solving for $\delta N^i$, and using $\frac{1}{N^2} = - g^{00}$,
$N^i = - g^{0i}/g^{00}$,
we eventually arrive at a formula for the variation of $N^i$ in terms
of only the metric and its variations:
\be
\delta N^i &=& - (\delta g^{0i} - \frac{g^{0i}}{g^{00}} \delta g^{00} )
( g^{00} + \delta g^{00})^{-1} \, .
\en
$g^{0i}$ has the transformation law
\be
g^{0i} &\rightarrow& g^{\alpha \beta} \left( \dd{(t+\pi)}{x^\alpha} \right)
\left( \dd{x^i}{x^\beta} \right) \nn\\
  &=& g^{0i} + g^{0i} \dot{\pi} + g^{ij} \d_j \pi \nn\\
\delta g^{0i} &=& \frac{N^i}{N^2} \dot{\pi}
  +  \hat g^{ij} \d_j \pi \, .
\en
We therefore find, to second order in the fields (which is all we need
to get the cubic action)
\be
\delta N^i &=& - (  \frac{N^i}{N^2} \dot{\pi}
 + \frac{\d^i \pi}{a^2} - N^i \delta g^{00} ) ( g^{00} + \delta g^{00})^{-1}
\nn\\
  &=&
    \d^i \pi + 2 \delta N \d^i \pi + 3 N^i \dot{\pi} - 2 \d^i \pi \dot{\pi} \, .
\en
The variation of $\hat g^{ij} \partial_t \hat g_{ij}$ also takes some work.
It is clear that the ``spatial'' components of the 4d metric $g^{ij}$ do not change under
time diffs, but
$g^{ij} = \hat g^{ij} - \frac{N^i N^j}{N^2} $, so $\hat g^{ij}$ transforms.
Using our transformation laws for $N^i$ and $1/N^2$, we find to second
order in the fields that
\be
\frac{N^i N^j}{N^2} &\rightarrow& \frac{N^i N^j}{N^2}
  + 2 \d^{(i} \pi N^{j)} + \d^i \pi \d^j \pi \, ,
\en
where round brackets stand for symmetrization. This implies
\be
\hat g^{ij} &\rightarrow& \hat g^{ij} + 2 \partial^{(i} \pi N^{j)} + \partial^i \pi
\partial^j \pi \, ,
\en
Incidentally, for terms at second
order in the fields, we can raise and lower spatial indices with
the unvaried $\hat g_{ij}$ and $\hat g^{ij}$, since corrections to this are cubic
order in the fields.  Also, we could have derived the above transformation
 of $\hat g_{ij}$ by calculating the transformation of the ``spatial'' components of the 4d
metric
$g_{ij}$.  It
is straightforward to see that both methods agree, since
\be
g_{ij} &\rightarrow& \dd{x^\alpha}{x'^i} \dd{x^\beta}{x'^j} g_{\alpha \beta}\, ,
\en
and $\dd{x^\alpha}{x'^i}$ is the $\alpha\;i$ component of the inverse of the matrix
\be
\dd{x'^\alpha}{x^\beta} &=& \bea{cc}
  1+ \dot{\pi} & \d_i \pi \\
   0 & \mathbf{1} \ena_{\alpha\beta} \, .
\en
Here Greek indexes run from 0 to 3 while Latin indexes run from 1 to 3.
Thus, to second order in the fields,
\be
\hat g^{ij} \partial_t \hat g_{ij} &\rightarrow&
  \frac{6 H}{1+\dot{\pi}} - \d_t (2 \d_i \pi N^i + (\d \pi)^2)\, ,
\en
and finally, we have
\be
\delta E_i^i &\rightarrow& \delta E_i^i - 3 H \dot{\pi}
     + 3 H \dot{\pi}^2 - \half \d_t ( 2 \d_i \pi N^i + (\d \pi)^2) \nn\\
  &-& \d^2 \pi - \d_i (2 \delta N \d^i \pi
  + 3 N^i \dot{\pi} - 2 \d^i \pi \dot{\pi})\, .
\en

\end{document}